\documentclass[aps,amssymb,onecolumn,amsmath,showpacs]{revtex4}

\usepackage{t1enc}
\usepackage[latin1]{inputenc}
\usepackage[english]{babel}
\usepackage[dvips]{color}
\usepackage{bm}
\usepackage{latexsym}
\usepackage{longtable}
\usepackage{graphicx}

\begin{document}
\preprint{GEF-TH-06/2006}
\author{Elena Santopinto}
\email{elena.santopinto@ge.infn.it}
\affiliation{INFN and Universit\`a di Genova,
via Dodecaneso 33, 16142 Genova, Italy}

\author{Giuseppe Galat\`a}
\affiliation{INFN and Universit\`a di Genova,
via Dodecaneso 33, 16142 Genova, Italy}
\title{Spectroscopy of tetraquark states.}

\begin{abstract}
A complete classification of $qq \bar q \bar q$ tetraquark states in terms of the spin-flavor, color and spatial degrees of freedom has been constructed. 
The permutation symmetry properties of both the spin-flavor and orbital parts of the $qq$ and $\bar{q}\bar{q}$ subsystems are discussed.
This complete classification is general and model independent and it is useful both for model builders and experimentalists. 
The total wave functions are also explicitly constructed in the hypothesis of ideal mixing; this basis for tetraquark states will enable the eigenvalue problem to be solved for a definite dynamical model.
An evaluation of the tetraquark spectrum is obtained from the Iachello mass formula
for normal mesons, here generalized to tetraquark systems. This mass formula is a generalization of the Gell-Mann Okubo mass 
formula, whose coefficients have been upgraded by a study of the latest PDG data. The ground state tetraquark nonet is identified with $f_{0}(600),\kappa(800), f_{0}(980), a_{0}(980)$.
The diquark-antidiquark limit is also studied. 
\end{abstract}

\pacs{14.40.Cs,12.39.-x, 02.20.-a }

\maketitle

\section{\label{sec:introduzione}Introduction}

The KLOE, E791 and BES collaborations have recently provided evidence of the low mass resonances $f_{0}(600)$ \cite{Aloisio:2002bt, Aitala:2000xu, Ablikim:2005ni}, formerly called $\sigma(450)$, and $\kappa (800)$ \cite{Aitala:2000xu, Ablikim:2005ni}, triggering new interest in meson spectroscopy. Maiani et al. \cite{Maiani:2004uc} have suggested that the lowest lying scalar mesons, $f_{0}(980)$, $a_{0}(980)$, $\kappa (800)$ and 
$f_0(600)$ could be described not as $q\bar q$ states, but as more complex tetraquark states, in particular as two clusters of two quarks and two antiquarks, i.e. a diquark and antidiquark system. The quark-antiquark assignment to P-waves \cite{Tornqvist:1995kr} has never really
 worked in the scalar case \cite{Jaffe:1976ig, Jaffe:1976ih}. Moreover, the $f_{0}(980)$ is more 
associated to strange than to up or down quarks as can be inferred from its higher mass and its decays \cite{Jaffe:1976ig, Jaffe:1976ih, Maiani:2004uc}, 
while in a simple quark-antiquark scheme it is associated with non-strange quarks \cite{Tornqvist:1995kr}; for this reason it is difficult to explain both its mass and its decay properties \cite{Jaffe:1976ig, Jaffe:1976ih, Maiani:2004uc}
 at the same time. One of the arguments by Jaffe \cite{Jaffe:1976ig, Jaffe:1976ih} and Maiani \cite{Maiani:2004uc} against the hypothesis of simple $q\bar q$ states is the observation that the experimental mass spectrum corresponding to this nonet is like a parabola with  a maximum in  the centre of the nonet corresponding to the $f_{0}(980)$ and $a_{0}(980)$, while in the $q\bar q$ case the parabola would be reversed and so the maximum would be at the edge of the nonet.

Other identifications have been proposed  \cite{Close:2002zu}, in particular 
quasi molecular-states (see Refs. \cite{Weinstein:1982gc,Weinstein:1983gd,Weinstein:1990gu} and references therein, \cite{Tornqvist:1995kr}) and uncorrelated $qq\bar q\bar q$ \cite{Jaffe:1978bu, Alford:2000mm}. 
Previous works on heavy tetraquark mesons can be found in Refs. \cite{Lipkin:1977ie,Black:1998wt,Pelaez:2003dy,Brink:1998as} and for light mesons in \cite{Brink:1994ic,Chan:1977st, Jaffe:2004ph} and references therein.  
As early as the 1970s Jaffe studied tetraquark systems in a bag model and discussed the resulting rich spectrum together with the problem of the missing resonances \cite{Jaffe:1976ig, Jaffe:1976ih, Jaffe:1978bu}. 
For review articles both on the experiments and on the theoretical models we refer the reader to \cite{Amsler:2004ps,Close:2002zu,PDG}.

In this article, we address the problem of constructing a complete classification scheme of the two quark-two antiquark states in terms of $SU_{sf}(6)$.
We identify the representations that contain exotics, i.e. states that cannot be constructed by $q\bar q$ only.
The tetraquark $O(3)\otimes SU_{sf}(6) \otimes SU_{c}(3)$
wave functions are explicitly constructed for the first time. They should be color singlets and, since they are composed of two quarks and two antiquarks, i. e. two couples of identical fermions, they should be antisymmetric for the exchange of the two quarks and the two antiquarks.  
The permutation symmetry properties of both the spin-flavor and the orbital parts of the $qq$ and $\bar{q}\bar{q}$
subsystems are discussed. The total wave functions are also explicitly constructed in the ideal mixing hypothesis, and can be useful in order to construct
tetraquark models. Finally, an evaluation of the tetraquark spectrum for the lowest scalar mesons is obtained from a generalization of the Iachello mass formula
for normal mesons \cite{Iachello:1991re}.

The classification of the states is general and is valid whichever dynamical model for tetraquarks is chosen. As an application, in section \ref{sec:diqantidiq} we develop a simple diquark-antidiquark model with no spatial excitations inside diquarks.  The states  are a subset of the general case.

\section{\label{sec:classificazionestati}The classification of tetraquark states}

As for all multiquark systems, the tetraquark wave function contains contributions connected to the spatial degrees of freedom and the internal degrees of freedom of color, flavor and spin.
In order to classify the corresponding states, we shall make use as much as possible of symmetry principles without, for the moment, introducing any explicit dynamical model.
In the construction of the classification scheme we are guided by two conditions: the tetraquark wave functions should be a color singlet, as all physical states, and since tetraquarks are composed of two couples of identical fermions, their states must be antisymmetric for the exchange of the two quarks and the two antiquarks.

In the following, we adopt the usual notation $[R]$ for the representations, where $R$ is the dimension of the representation.

\subsection{\label{subsec:sapore}The SU(3)$_{f}$-flavor classification of $qq\bar q\bar q$ states}

The allowed SU(3)$_{f}$ representations for the $qq\bar q\bar q$ mesons are obtained by means of the product
\begin{equation}
[3]\otimes [\bar 3]\otimes [3]\otimes [\bar 3] =[1]\oplus [8]\oplus [8]\oplus [10]\oplus [8]\oplus [8]\oplus [1]\oplus [\overline{10}]\oplus [27]
\label{eq:saporetetraquark}
\end{equation}
The allowed isospin values are $I=0,\frac{1}{2},1,\frac{3}{2},2$ , while the hypercharge values are $Y=0,\pm 1,\pm 2$.
We can notice that the values $I=\frac{3}{2},2$ and $Y=\pm 2$ are exotic, which means that they are forbidden for the $q\bar q$ mesons.

In Appendix \ref{app:statisapore} the flavor states in the $qq\bar q\bar q$ configuration are explicitly written.

\subsection{\label{subsec:colore}The SU(3)$_{c}$-color classification of $qq\bar q\bar q$ states}
Color representations for $qq\bar q\bar q$ mesons are those written in (\ref{eq:saporetetraquark}) for the flavor case. 
However, the only color representation allowed for mesons (or in general for any isolated particle) is the singlet, so there are two colour representations for $qq\bar q\bar q$ mesons, while there is only one singlet for normal mesons. This fact implies that color for tetraquarks is not a trivial quantum number as it was for conventional mesons.

\subsection{\label{subsec:spin}The SU(2)$_{s}$-spin classification of $qq\bar q\bar q$ states}
The $qq\bar q\bar q$ spin states are given by the product
\begin{equation}
[2]\otimes [2]\otimes [2]\otimes [2] =[1]\oplus [3]\oplus [1]\oplus [3]\oplus [3]\oplus [5]
\end{equation} 
We can see that tetraquarks can have an exotic spin $S=2$, value forbidden for $q\bar q$ mesons.

In Appendix \ref{app:statispin} the spin states in the $qq\bar q\bar q$ configuration are explicitly written.

\subsection{\label{subsec:spinsapore}The SU(6)$_{sf}$-spin-flavor classification of $qq\bar q\bar q$ states}
The spin-flavor SU(6)$_{sf}$ $qq\bar q\bar q$ representations are obtained by means of the product
\begin{equation}
[6]\otimes [\bar 6]\otimes [6]\otimes [\bar 6]=[1]\oplus [35]\oplus [35]\oplus [405]\oplus [35]\oplus [280]\oplus [35]\oplus [1]\oplus [\overline{280}]\oplus [189]
\end{equation}

A complete classification of the tetraquark states involves the analysis of the flavor and spin content of each spin-flavor representation , i.e. the decomposition of the representation of SU(6)$_{sf}$ into those of SU(3)$_{f}\otimes $SU(2)$_{s}$ in the notation $[\text{flavor repr.,\; spin repr.}]$, 
\begin{equation}
[189]=[8,5]\oplus [10,3]\oplus [27,1]\oplus [\overline{10},3]\oplus 2[8,3]\oplus [8,1]\oplus [1,1]\oplus [1,5]
\end{equation}
\begin{equation}
[280]=[10,5]\oplus [8,5]\oplus [27,3]\oplus [10,3]\oplus 2[8,3]\oplus [10,1]\oplus [\overline{10},1]\oplus [8,1]\oplus [1,3]
\end{equation}
\begin{equation}
[\overline{280}]=[\overline{10},5]\oplus [8,5]\oplus [27,3]\oplus [\overline{10},3]\oplus 2[8,3]\oplus [\overline{10},1]\oplus [10,1]\oplus [8,1]\oplus [1,3]
\end{equation}
\begin{equation}
[405]=[1,1]\oplus [1,5]\oplus [8,5]\oplus 2[8,3]\oplus [27,1]\oplus [8,1]\oplus [27,3]\oplus [10,3]\oplus [\overline{10},3]\oplus [27,5]
\end{equation}
\begin{equation}
[1]=[1,1]
\end{equation}
\begin{equation}
[35]=[1,3]\oplus [8,1]\oplus [8,3]
\end{equation}

\subsection{\label{subsec:JPC}Angular momentum, parity and charge conjugation quantum numbers}
The total angular momentum, parity and charge conjugation quantum numbers for the $q\bar q$ mesons are well known. Thus, here we recall only that the following $J^{PC}$ combinations are forbidden for normal mesons:
\begin{equation}
0^{--},(even)^{+-},(odd)^{-+}.
\end{equation}

\begin{figure}
\caption{\label{fig:Jtetraquark2} $qq\bar q\bar q$ mesons' total angular momentum scheme}
\includegraphics[width=13cm]{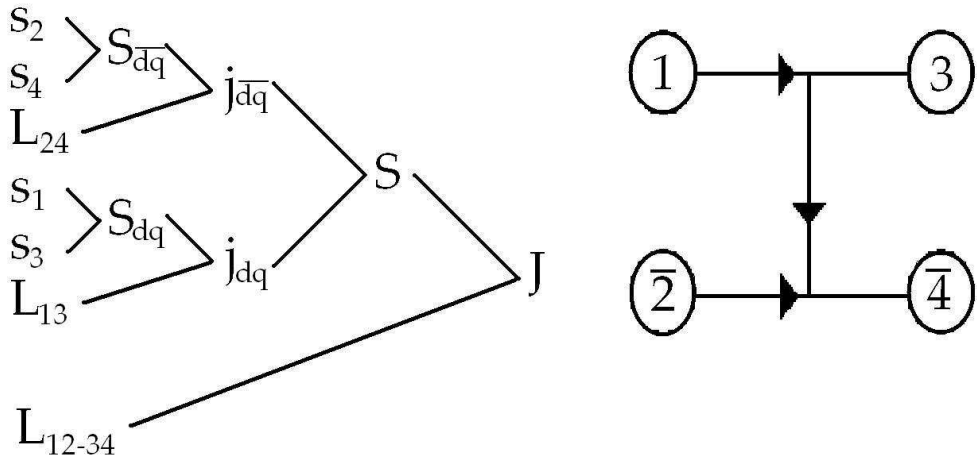}  
\end{figure}

Tetraquarks are made up of four objects, so we have to define three relative coordinates (see figure \ref{fig:Jtetraquark2}) \cite{Estrada}
\begin{subequations}
\begin{eqnarray}
& \vec r_{13}=\vec r_{3}-\vec r_{1} & \\
& \vec r_{24}=\vec r_{4}-\vec r_{2} & \\
& \vec r_{12-34}=\vec r_{CM24}-\vec r_{CM13}=\frac{m_{2}\vec r_{2}+m_{4}\vec r_{4}}{m_{2}+m_{4}}-\frac{m_{1}\vec r_{1}+m_{3}\vec r_{3}}{m_{1}+m_{3}}. &
\end{eqnarray} 
\end{subequations}
This is only a possible choice of coordinates. Other types of coordinates, useful to describe the strong decays, can be defined \cite{Brink:1994ic}.

In the tetraquark case, we have four different spins and three orbital angular momenta. The total angular momentum $J$ can be obtained by combining spins and orbital momenta, as shown in figure \ref{fig:Jtetraquark2}. 

The parity for a tetraquark system is the product of the intrinsic parities of the quarks and the antiquarks times the factors coming from the spherical harmonics  \cite{Estrada}.
\begin{equation}
\label{eq:paritàtetraquark}
P=P_{q}P_{q}P_{\bar q}P_{\bar q}(-1)^{L_{13}}(-1)^{L_{24}}(-1)^{L_{12-34}}=(-1)^{L_{13}+L_{24}+L_{12-34}}
\end{equation}

Using our coordinates, tetraquark charge conjugation eigenvalues can be calculated by following the same steps as in the $q\bar q$ case. Indeed, we can consider a tetraquark as a $Q\bar Q$ meson, where $Q$ represents the couple of quarks and $\bar Q$ the couple of antiquarks (see figure \ref{fig:Jtetraquark2}), with total \textquotedblleft spin\textquotedblright $S$ and relative angular momentum $L_{12-34}$. The $C$ eigenvectors are those states for which $Q$ and $\bar Q$ have opposite charges. So applying the charge conjugation operator to these mesons is the same as exchanging the couple of quarks with the couple of antiquarks. The factors arising from this exchange are the $C$ operator eigenvalues \cite{Estrada}:
\begin{equation}
\label{eq:concaricatetraquark}
C=(-1)^{L_{12-34}+S}
\end{equation}

Tetraquark mesons do not have forbidden $J^{PC}$ combinations because they have more degrees of freedom (in particular they have three different orbital angular momenta) than the normal mesons. 

\subsection{\label{subsec:statiglobali}Tetraquark states and the Pauli principle}
Tetraquarks are composed of two couples of identical fermions, so their states must be antisymmetric for the exchange of the two quarks and the two antiquarks. 
In this respect it is necessary to study the permutation symmetry (i.e. the irreducible representations of S$_{2}$) of the color, flavor, spin and spatial parts of the wave functions of each subsystem, two quarks and two antiquarks.

Only the singlet color states are physical states, so there are only two color singlets and we write them by underlining their permutation S$_{2}$ symmetry, which can be only antisymmetric (A) or symmetric (S): 
\begin{subequations}
\begin{eqnarray}
 & (qq)\; \text{in} \;[\bar 3]_{C}\; \text{(A)\; and} \;(\bar q\bar q)\; \text{in}\; [3]_{C}\; \text{(A)}, & \\
 & (qq)\; \text{in}\; [6]_{C} \;\text{(S)\; and}\; (\bar q\bar q)\; \text{in}\; [\bar 6]_{C}\; \text{(S)}. &
\end{eqnarray} 
\end{subequations} 

Next, we study the permutation symmetry of the spatial part of the two quarks (two antiquarks) states.
The permutation symmetry of the spatial part of the couple of quarks and antiquarks is
\begin{subequations}
\begin{eqnarray}
& (qq)\; \text{with}\; L_{13}\; \text{even\; (S)},\; (\bar q\bar q)\; \text{with}\; L_{24}\; \text{even\; (S)} & \\
& (qq)\; \text{with}\; L_{13}\; \text{odd\; (A)},\; (\bar q\bar q)\; \text{with}\; L_{24}\; \text{odd\; (A)} & \\
& (qq)\; \text{with}\; L_{13}\; \text{even\; (S)},\; (\bar q\bar q)\; \text{with}\; L_{24}\; \text{odd\; (A)} & \\
& (qq)\; \text{with}\; L_{13}\; \text{odd\; (A)},\; (\bar q\bar q)\; \text{with}\; L_{24}\; \text{even\; (S)} &
\end{eqnarray} 
\end{subequations} 
The permutation symmetry of the spatial part derives from the parity of the couple of quarks and antiquarks, which are respectively $P_{qq}=P_{q}P_{q}(-1)^{L_{13}}=(-1)^{L_{13}}$ and $P_{\bar q\bar q}=P_{\bar q}P_{\bar q}(-1)^{L_{24}}=(-1)^{L_{24}}$. 

The permutation symmetry of the SU(6)$_{sf}$ representations for a couple of quarks is written below.
\begin{subequations}
\begin{eqnarray}
& [15]_{sf}\; \text{(A),\; which\; means\; symmetric\; spin}\; (S_{dq}=1)\; \text{and\; antisymmetric\; flavor\;} ([\bar 3]_{f})\;  & \nonumber \\
& \text{or\;  antisymmetric\; spin}\; (S_{dq}=0)\; \text{and\; symmetric\; flavor}\; ([6]_{f}) & \\
& [21]_{sf}\; \text{(S),\; which\; means\; symmetric\; spin}\; (S_{dq}=1)\; \text{and\; symmetric\; flavor} \; ([6]_{f})  & \nonumber \\
& \text{or\; antisymmetric\; spin}\; (S_{dq}=0)\; \text{and\; antisymmetric\; flavor}\; ([\bar 3]_{f}) &
\end{eqnarray} 
\end{subequations} 
The spin-flavor representations for the couple of antiquarks are the conjugate representations $[\overline{15}]_{sf}$ (A) and $[\overline{21}]_{sf}$ (S). 

The spatial, flavor, color and spin parts with given permutation symmetry (S$_{2}$) must now be arranged together to obtain completely antisymmetric states under the exchange of the two quarks and the two antiquarks. The resulting states are listed in Table \ref{tab:statiglobali}.
\begin{table*}
\begin{center}
\caption{\label{tab:statiglobali}Allowed color, flavor and spin tetraquark states.}
\begin{tabular}{|c|c|c|c|c|c|c|c|}
\hline
color $(qq)\otimes (\bar q\bar q)$ & $L_{13}$ & $L_{24}$ & $S_{dq}$ & $S_{d\bar q}$ & flavor $(qq)\otimes (\bar q\bar q)$ & $S_{tot}$ & total flavor \\
\hline
$[\bar 3]_{c}\otimes [3]_{c}$ & even & even & 1 & 1 & $[6]_{f}\otimes [\bar 6]_{f}$ & 0,1,2 & $[1]\oplus [8]\oplus [27]$  \\
\hline
$[\bar 3]_{c}\otimes [3]_{c}$ & even & even & 1 & 0 & $[6]_{f}\otimes [3]_{f}$ & 1 & $[8]\oplus [10]$ \\
\hline
$[\bar 3]_{c}\otimes [3]_{c}$ & even & even & 0 & 1 & $[\bar 3]_{f}\otimes [\bar 6]_{f}$ & 1 & $[8]\oplus [\overline{10}]$ \\
\hline
$[\bar 3]_{c}\otimes [3]_{c}$ & even & even & 0 & 0 & $[\bar 3]_{f}\otimes [3]_{f}$ & 0 & $[1]\oplus [8]$ \\
\hline
$[6]_{c}\otimes [\bar 6]_{c}$ & even & even & 1 & 1 & $[\bar 3]_{f}\otimes [3]_{f}$ & 0,1,2 & $[1]\oplus [8]$  \\
\hline
$[6]_{c}\otimes [\bar 6]_{c}$ & even & even & 1 & 0 & $[\bar 3]_{f}\otimes [\bar 6]_{f}$ & 1 & $[8]\oplus [\overline{10}]$ \\
\hline
$[6]_{c}\otimes [\bar 6]_{c}$ & even & even & 0 & 1 & $[6]_{f}\otimes [3]_{f}$ & 1 & $[8]\oplus [10]$ \\
\hline
$[6]_{c}\otimes [\bar 6]_{c}$ & even & even & 0 & 0 & $[6]_{f}\otimes [\bar 6]_{f}$ & 0 & $[1]\oplus [8]\oplus [27]$ \\
\hline
$[\bar 3]_{c}\otimes [3]_{c}$ & odd & odd & 1 & 1 & $[\bar 3]_{f}\otimes [3]_{f}$ & 0,1,2 & $[1]\oplus [8]$  \\
\hline
$[\bar 3]_{c}\otimes [3]_{c}$ & odd & odd & 1 & 0 & $[\bar 3]_{f}\otimes [\bar 6]_{f}$ & 1 & $[8]\oplus [\overline{10}]$ \\
\hline
$[\bar 3]_{c}\otimes [3]_{c}$ & odd & odd & 0 & 1 & $[6]_{f}\otimes [3]_{f}$ & 1 & $[8]\oplus [10]$ \\
\hline
$[\bar 3]_{c}\otimes [3]_{c}$ & odd & odd & 0 & 0 & $[6]_{f}\otimes [\bar 6]_{f}$ & 0 & $[1]\oplus [8]\oplus [27]$ \\
\hline
$[6]_{c}\otimes [\bar 6]_{c}$ & odd & odd & 1 & 1 & $[6]_{f}\otimes [\bar 6]_{f}$ & 0,1,2 & $[1]\oplus [8]\oplus [27]$  \\
\hline
$[6]_{c}\otimes [\bar 6]_{c}$ & odd & odd & 1 & 0 & $[6]_{f}\otimes [3]_{f}$ & 1 & $[8]\oplus [10]$ \\
\hline
$[6]_{c}\otimes [\bar 6]_{c}$ & odd & odd & 0 & 1 & $[\bar 3]_{f}\otimes [\bar 6]_{f}$ & 1 & $[8]\oplus [\overline{10}]$ \\
\hline
$[6]_{c}\otimes [\bar 6]_{c}$ & odd & odd & 0 & 0 & $[\bar 3]_{f}\otimes [3]_{f}$ & 0 & $[1]\oplus [8]$ \\
\hline
$[\bar 3]_{c}\otimes [3]_{c}$ & even & odd & 1 & 1 & $[6]_{f}\otimes [3]_{f}$ & 0,1,2 & $[8]\oplus [10]$  \\
\hline
$[\bar 3]_{c}\otimes [3]_{c}$ & even & odd & 1 & 0 & $[6]_{f}\otimes [\bar 6]_{f}$ & 1 & $[1]\oplus [8]\oplus [27]$ \\
\hline
$[\bar 3]_{c}\otimes [3]_{c}$ & even & odd & 0 & 1 & $[\bar 3]_{f}\otimes [\bar 6]_{f}$ & 1 & $[8]\oplus [\overline{10}]$ \\
\hline
$[\bar 3]_{c}\otimes [3]_{c}$ & even & odd & 0 & 0 & $[\bar 3]_{f}\otimes [3]_{f}$ & 0 & $[1]\oplus [8]$ \\
\hline
$[6]_{c}\otimes [\bar 6]_{c}$ & even & odd & 1 & 1 & $[\bar 3]_{f}\otimes [\bar 6]_{f}$ & 0,1,2 & $[8]\oplus [\overline{10}]$  \\
\hline
$[6]_{c}\otimes [\bar 6]_{c}$ & even & odd & 1 & 0 & $[\bar 3]_{f}\otimes [3]_{f}$ & 1 & $[1]\oplus [8]$ \\
\hline
$[6]_{c}\otimes [\bar 6]_{c}$ & even & odd & 0 & 1 & $[6]_{f}\otimes [\bar 6]_{f}$ & 1 & $[1]\oplus [8]\oplus [27]$ \\
\hline
$[6]_{c}\otimes [\bar 6]_{c}$ & even & odd & 0 & 0 & $[6]_{f}\otimes [3]_{f}$ & 0 & $[8]\oplus [10]$ \\
\hline
$[\bar 3]_{c}\otimes [3]_{c}$ & odd & even & 1 & 1 & $[\bar 3]_{f}\otimes [\bar 6]_{f}$ & 0,1,2 & $[8]\oplus [\overline{10}]$  \\
\hline
$[\bar 3]_{c}\otimes [3]_{c}$ & odd & even & 1 & 0 & $[\bar 3]_{f}\otimes [3]_{f}$ & 1 & $[1]\oplus [8]$ \\
\hline
$[\bar 3]_{c}\otimes [3]_{c}$ & odd & even & 0 & 1 & $[6]_{f}\otimes [\bar 6]_{f}$ & 1 & $[1]\oplus [8]\oplus [27]$ \\
\hline
$[\bar 3]_{c}\otimes [3]_{c}$ & odd & even & 0 & 0 & $[6]_{f}\otimes [3]_{f}$ & 0 & $[8]\oplus [10]$ \\
\hline
$[6]_{c}\otimes [\bar 6]_{c}$ & odd & even & 1 & 1 & $[6]_{f}\otimes [3]_{f}$ & 0,1,2 & $[8]\oplus [10]$  \\
\hline
$[6]_{c}\otimes [\bar 6]_{c}$ & odd & even & 1 & 0 & $[6]_{f}\otimes [\bar 6]_{f}$ & 1 & $[1]\oplus [8]\oplus [27]$ \\
\hline
$[6]_{c}\otimes [\bar 6]_{c}$ & odd & even & 0 & 1 & $[\bar 3]_{f}\otimes [\bar 6]_{f}$ & 1 & $[8]\oplus [\overline{10}]$ \\
\hline
$[6]_{c}\otimes [\bar 6]_{c}$ & odd & even & 0 & 0 & $[\bar 3]_{f}\otimes [3]_{f}$ & 0 & $[1]\oplus [8]$ \\
\hline
\end{tabular}
\end{center}
\end{table*}    
In this Table we write the color, flavor and spin of the couples of quarks and antiquarks and the corresponding total spin and flavor of the tetraquark states. The total color has been omitted since it is always a singlet.

We want, then, to determine the $J^{PC}$ (where $C$ is obviously intended only for its eigenstates) possible quantum numbers for a tetraquark with a given flavor and spin. The total angular momentum $J$ depends on the values of the three orbital angular momenta $L_{13}$, $L_{24}$ and $L_{12-34}$. For obvious reasons, we have chosen to study only the lower value cases, in particular only up to the case that at most one of the three angular momenta is one.
In Table \ref{tab:momang} we combine the orbital angular momenta with the spins to obtain the total angular momentum. 
\begin{table}
\begin{center}
\caption{\label{tab:momang}Tetraquark total angular momenta}
\begin{tabular}{|c|c|c|c|c|c|c|c|}
\hline
$L_{13}$ & $L_{24}$ & $L_{12-34}$ & $S_{dq}$ & $S_{d\bar q}$ & $j_{dq}$ & $j_{d\bar q}$ & $J$ \\
\hline
0 & 0 & 0 & 0 & 0 & 0 & 0 & 0 \\
\hline
0 & 0 & 0 & 0 & 1 & 0 & 1 & 1 \\
\hline
0 & 0 & 0 & 1 & 0 & 1 & 0 & 1 \\
\hline
0 & 0 & 0 & 1 & 1 & 1 & 1 & 0,1,2 \\
\hline
0 & 1 & 0 & 0 & 0 & 0 & 1 & 1 \\
\hline
0 & 1 & 0 & 0 & 1 & 0 & 0,1,2 & 0,1,2 \\
\hline
0 & 1 & 0 & 1 & 0 & 1 & 1 & 0,1,2 \\
\hline
0 & 1 & 0 & 1 & 1 & 1 & 0,1,2 & 0,1,2,3 \\
\hline
1 & 0 & 0 & 0 & 0 & 1 & 0 & 1 \\
\hline
1 & 0 & 0 & 0 & 1 & 1 & 1 & 0,1,2 \\
\hline
1 & 0 & 0 & 1 & 0 & 0,1,2 & 0 & 0,1,2 \\
\hline
1 & 0 & 0 & 1 & 1 & 0,1,2 & 1 & 0,1,2,3 \\
\hline
0 & 0 & 1 & 0 & 0 & 0 & 0 & 1 \\
\hline
0 & 0 & 1 & 0 & 1 & 0 & 1 & 0,1,2 \\
\hline
0 & 0 & 1 & 1 & 0 & 1 & 0 & 0,1,2 \\
\hline
0 & 0 & 1 & 1 & 1 & 1 & 1 & 0,1,2,3 \\
\hline
\end{tabular}
\end{center}
\end{table}
In Tables \ref{tab:JPC1}, \ref{tab:JPC2}, \ref{tab:JPC3} and \ref{tab:JPC4} we write the possible $J^{PC}$ combinations for every tetraquark with a given flavor and spin. 

\begin{table*}
\begin{center}
\caption{\label{tab:JPC1}color, flavor, spin ($S_{tot}$) and $J^{PC}$ for tetraquarks with $L_{13}=L_{24}=L_{12-34}=0$}
\begin{tabular}{|c|c|c|c|c|c|c|c|c|}
\hline
color & flavor & $S_{dq}$ & $S_{d\bar q}$ & $j_{dq}$ & $j_{d\bar q}$ & $S_{tot}$ & $S$ & $J^{PC}$  \\
\hline
$[\bar 3]\otimes [3]$ & $[1]\oplus [8]$ & 0 & 0 & 0 & 0 & 0 & 0 & $0^{++}$   \\
\hline
$[\bar 3]\otimes [3]$ & $[8]\oplus [\overline{10}]$ & 0 & 1 & 0 & 1 & 1 & 1 &  $1^{+}$  \\
\hline
$[\bar 3]\otimes [3]$ & $[8]\oplus [10]$ & 1 & 0 & 1 & 0 & 1 & 1 & $1^{+}$   \\
\hline
$[\bar 3]\otimes [3]$ & $[1]\oplus [8]\oplus [27]$ & 1 & 1 & 1 & 1 & 0 & 0 & $0^{++}$   \\
                      &                            &   &   &   &   & 1 & 1 & $1^{+-}$   \\
                      &                            &   &   &   &   & 2 & 2 & $2^{++}$   \\
\hline
$[6]\otimes [\bar 6]$ & $[1]\oplus [8]\oplus [27]$ & 0 & 0 & 0 & 0 & 0 & 0 & $0^{++}$   \\
\hline
$[6]\otimes [\bar 6]$ & $[8]\oplus [10]$ & 0 & 1 & 0 & 1 & 1 & 1 &  $1^{+}$  \\
\hline
$[6]\otimes [\bar 6]$ & $[8]\oplus [\overline{10}]$ & 1 & 0 & 1 & 0 & 1 & 1 & $1^{+}$   \\
\hline
$[6]\otimes [\bar 6]$ & $[1]\oplus [8]$ & 1 & 1 & 1 & 1 & 0 & 0 & $0^{++}$   \\
                      &                 &   &   &   &   & 1 & 1 & $1^{+-}$   \\
                      &                 &   &   &   &   & 2 & 2 & $2^{++}$   \\
\hline
\end{tabular}
\end{center}
\end{table*}

\begin{table*}
\begin{center}
\caption{\label{tab:JPC2}color, flavor, spin ($S_{tot}$) and $J^{PC}$ for tetraquarks with $L_{13}=L_{12-34}=0$ and $L_{24}=1$}
\begin{tabular}{|c|c|c|c|c|c|c|c|c|}
\hline
color & flavor & $S_{dq}$ & $S_{d\bar q}$ & $j_{dq}$ & $j_{d\bar q}$ & $S_{tot}$ & $S$ & $J^{PC}$  \\
\hline
$[\bar 3]\otimes [3]$ & $[1]\oplus [8]$ & 0 & 0 & 0 & 1 & 0 & 1 & $1^{-}$   \\
\hline
$[\bar 3]\otimes [3]$ & $[8]\oplus [\overline{10}]$ & 0 & 1 & 0 & 0  & 1 & 0 & $0^{-}$  \\
                      &                             &   &   &   & 1  & 1 & 1 & $1^{-}$   \\
                      &                             &   &   &   & 2  & 1 & 2 & $2^{-}$   \\
\hline
$[\bar 3]\otimes [3]$ & $[1]\oplus [8]\oplus [27]$ & 1 & 0 & 1 & 1 & 1 & 0 & $0^{-+}$   \\
                      &                            &   &   &   &   &   & 1 & $1^{--}$   \\
                      &                            &   &   &   &   &   & 2 & $2^{-+}$   \\
\hline
$[\bar 3]\otimes [3]$ & $[8]\oplus [10]$           & 1 & 1 & 1 & 0 & 0,1,2 & 1 & $1^{-}$   \\
                      &                            &   &   &   & 1 & 0,1,2  & 0,1,2 & $0^{-},1^{-},2^{-}$   \\
                      &                            &   &   &   & 2 & 0,1,2  & 1,2,3 & $1^{-},2^{-},3^{-}$   \\
\hline
$[6]\otimes [\bar 6]$ & $[8]\oplus [10]$           & 0 & 0 & 0 & 1 & 0 & 1 & $1^{-}$   \\
\hline
$[6]\otimes [\bar 6]$ & $[1]\oplus [8]\oplus [27]$ & 0 & 1 & 0 & 0  & 1 & 0 & $0^{-+}$  \\
                      &                            &   &   &   & 1  & 1 & 1 & $1^{-}$   \\
                      &                            &   &   &   & 2  & 1 & 2 & $2^{-}$   \\
\hline
$[6]\otimes [\bar 6]$ & $[1]\oplus [8]$            & 1 & 0 & 1 & 1 & 1 & 0 & $0^{-+}$   \\
                      &                            &   &   &   &   &   & 1 & $1^{--}$   \\
                      &                            &   &   &   &   &   & 2 & $2^{-+}$   \\
\hline
$[6]\otimes [\bar 6]$ & $[8]\oplus [\overline{10}]$ & 1 & 1 & 1 & 0 & 0,1,2 & 1 & $1^{-}$   \\
                      &                             &   &   &   & 1 & 0,1,2  & 0,1,2 & $0^{-},1^{-},2^{-}$   \\
                      &                             &   &   &   & 2 & 0,1,2  & 1,2,3 & $1^{-},2^{-},3^{-}$   \\
\hline
\end{tabular}
\end{center}
\end{table*}

\begin{table*}
\begin{center}
\caption{\label{tab:JPC3}color, flavor, spin ($S_{tot}$) and $J^{PC}$ for tetraquarks with $L_{24}=L_{12-34}=0$ and $L_{13}=1$}
\begin{tabular}{|c|c|c|c|c|c|c|c|c|}
\hline
color & flavor & $S_{dq}$ & $S_{d\bar q}$ & $j_{dq}$ & $j_{d\bar q}$ & $S_{tot}$ & $S$ & $J^{PC}$  \\
\hline
$[\bar 3]\otimes [3]$ & $[8]\oplus [10]$ & 0 & 0 & 1 & 0 & 0 & 1 & $1^{-}$   \\
\hline
$[\bar 3]\otimes [3]$ & $[1]\oplus [8]\oplus [27]$  & 0 & 1 & 1 & 1  & 1 & 0 & $0^{-+}$  \\
                      &                             &   &   &   &    &   & 1 & $1^{--}$   \\
                      &                             &   &   &   &    &   & 2 & $2^{-+}$   \\
\hline
$[\bar 3]\otimes [3]$ & $[1]\oplus [8]$            & 1 & 0 & 0 & 0 & 1 & 0 & $0^{-+}$   \\
                      &                            &   &   & 1 & 0 & 1 & 1 & $1^{-}$   \\
                      &                            &   &   & 2 & 0 & 1 & 2 & $2^{-}$   \\
\hline
$[\bar 3]\otimes [3]$ & $[8]\oplus [\overline{10}]$           & 1 & 1 & 0 & 1 & 0,1,2 & 1 & $1^{-}$   \\
                      &                            &   &   & 1 & 1 & 0,1,2  & 0,1,2 & $0^{-},1^{-},2^{-}$   \\
                      &                            &   &   & 2 & 1 & 0,1,2  & 1,2,3 & $1^{-},2^{-},3^{-}$   \\
\hline
$[6]\otimes [\bar 6]$ & $[1]\oplus [8]$           & 0 & 0 & 1 & 0 & 0 & 1 & $1^{-}$   \\
\hline
$[6]\otimes [\bar 6]$ & $[8]\oplus [\overline{10}]$ & 0 & 1 & 1 & 1  & 1 & 0 & $0^{-}$  \\
                      &                             &   &   &   &    &   & 1 & $1^{-}$   \\
                      &                             &   &   &   &    &   & 2 & $2^{-}$   \\
\hline
$[6]\otimes [\bar 6]$ & $[1]\oplus [8]\oplus [27]$ & 1 & 0 & 0 & 0 & 1 & 0 & $0^{-+}$   \\
                      &                            &   &   & 1 & 0 & 1 & 1 & $1^{-}$   \\
                      &                            &   &   & 2 & 0 & 1 & 2 & $2^{-}$   \\
\hline
$[6]\otimes [\bar 6]$ & $[8]\oplus [10]$ & 1 & 1 & 0 & 1 & 0,1,2 & 1 & $1^{-}$   \\
                      &                            &   &   & 1 & 1 & 0,1,2  & 0,1,2 & $0^{-},1^{-},2^{-}$   \\
                      &                            &   &   & 2 & 1 & 0,1,2  & 1,2,3 & $1^{-},2^{-},3^{-}$   \\
\hline
\end{tabular}
\end{center}
\end{table*}

\begin{table*}
\begin{center}
\caption{\label{tab:JPC4}color, flavor, spin ($S_{tot}$) and $J^{PC}$ for tetraquarks with $L_{24}=L_{13}=0$ and $L_{12-34}=1$}
\begin{tabular}{|c|c|c|c|c|c|c|c|c|}
\hline
color & flavor & $S_{dq}$ & $S_{d\bar q}$ & $j_{dq}$ & $j_{d\bar q}$ & $S_{tot}$ & $S$ & $J^{PC}$  \\
\hline
$[\bar 3]\otimes [3]$ & $[1]\oplus [8]$ & 0 & 0 & 0 & 0 & 0 & 0 & $1^{--}$   \\
\hline
$[\bar 3]\otimes [3]$ & $[8]\oplus [\overline{10}]$  & 0 & 1 & 0 & 1  & 1 & 1 & $0^{-}$  \\
                      &                             &   &   &   &    &   &   & $1^{-}$   \\
                      &                             &   &   &   &    &   &   & $2^{-}$   \\
\hline
$[\bar 3]\otimes [3]$ & $[8]\oplus [10]$            & 1 & 0 & 1 & 0 & 1 & 1 & $0^{-}$   \\
                      &                            &   &   &   &   &   &   & $1^{-}$   \\
                      &                            &   &   &   &   &   &   & $2^{-}$   \\
\hline
$[\bar 3]\otimes [3]$ & $[1]\oplus [8]\oplus [27]$ & 1 & 1 & 1 & 1 & 0  & 0 & $1^{--}$   \\
                      &                            &   &   &   &   & 1  & 1 & $0^{-+},1^{-+},2^{-+}$   \\
                      &                            &   &   &   &   & 2  & 2 & $1^{--},2^{-+},3^{--}$   \\
\hline
$[6]\otimes [\bar 6]$ & $[1]\oplus [8]\oplus [27]$           & 0 & 0 & 0 & 0 & 0 & 0 & $1^{--}$   \\
\hline
$[6]\otimes [\bar 6]$ & $[8]\oplus [10]$ & 0 & 1 & 0 & 1  & 1 & 1 & $0^{-}$  \\
                      &                             &   &   &   &    &   &   & $1^{-}$   \\
                      &                             &   &   &   &    &   &   & $2^{-}$   \\
\hline
$[6]\otimes [\bar 6]$ & $[8]\oplus [\overline{10}]$            & 1 & 0 & 1 & 0 & 1 & 1 & $0^{-}$   \\
                      &                            &   &   &   &   &   &   & $1^{-}$   \\
                      &                            &   &   &   &   &   &   & $2^{-}$   \\
\hline
$[6]\otimes [\bar 6]$ & $[1]\oplus [8]$ & 1 & 1 & 1 & 1 & 0  & 0 & $1^{--}$   \\
                      &                            &   &   &   &   & 1  & 1 & $0^{-+},1^{-+},2^{-+}$   \\
                      &                            &   &   &   &   & 2  & 2 & $1^{--},2^{-+},3^{--}$   \\
\hline
\end{tabular}
\end{center}
\end{table*}

\subsection{\label{subsec:paritàG}G parity}

Charged particles are not eigenstates of $C$ since $C$ takes a positive particle into a negative particle and vice versa. 
$G$ parity is a generalization of the concept of $C$ parity such that members of an isospin multiplet can each be assigned a good quantum number that would reproduce $C$ for the neutral particle.
The $G$ operator is defined as the combination of $C$ and a $\pi $ rotation around the $y$ axis in the isospin space,
\begin{equation}
G=C\mathcal{R}_{y}(\pi )=Ce^{i\pi I_{2}}.
\end{equation}

The $G$ eigenstates are tetraquark states with flavor charges equal to zero, i.e. strangeness equal to zero in the light mesons case, and their eigenvalues are:
\begin{equation}
\label{eq:paritàG}
G=(-1)^{L_{12-34}+S+I}.
\end{equation}
The states belonging to $[8]\oplus [10]$ and $[8]\oplus [\overline{10}]$ flavor multiplets are the only exceptions to the validity of Equation (\ref{eq:paritàG}). 
Actually a linear combination \cite{Jaffe:1977cv} of these states diagonalizes the $G$ parity.
\begin{subequations}
\begin{eqnarray}
  |\psi ^{+}_{G}> & = & \frac{1}{\sqrt{2}}(|[8]\oplus [10]>+|[8]\oplus [\overline{10}]>) \\
  |\psi ^{-}_{G}> & = & \frac{1}{\sqrt{2}}(|[8]\oplus [10]>-|[8]\oplus [\overline{10}]>) 
\end{eqnarray}
\end{subequations}
where $|\psi ^{+}_{G}>$ and $|\psi ^{-}_{G}>$ are the $G$ parity eigenvectors with eigenvalues  $G=(-1)^{L_{12-34}+S+I+1}$ and $G=(-1)^{L_{12-34}+S+I}$ respectively.

\section{\label{sec:spettro}The Iachello, Mukhopadhyay and Zhang mass formula for $q\bar q$ mesons.}

In 1991 Iachello, Mukhopadhyay and Zhang developed a mass formula \cite{Iachello:1991re,Iachello:1991fj} for $q\bar q$ mesons, which is a generalization of the G\"{u}rsey and Radicati mass formula \cite{Gursey:1992dc,Gursey:1964ic},
\begin{equation}
M^{2}=(N_{n}M_{n}+N_{s}M_{s})^{2}+a\cdot \nu +b\cdot L+c\cdot S+d\cdot J+h\cdot <M'^{2}>_{ij,i'j'}+i\cdot <M''^{2}>_{ij,i'j'},
\label{eq:formulamassa}
\end{equation}
where $N_{n}$ is the non-strange quark and antiquark number, $M_{n}=M_{u}=M_{d}$ is the non-strange constituent quark mass, $N_{s}$ is the strange quark and antiquark number, $M_{s}$ is the strange constituent quark mass, $\nu $ is the vibrational quantum number, $L$ is the orbital angular momentum, $S$ the total spin and $J$ the total angular momentum. $<M'^{2}>_{ij,i'j'}$ and $<M'^{2}>_{ij,i'j'}$ are two phenomenological terms which act only on the lowest pseudoscalar mesons. Specifically, the first acts on the octet; it encodes the unusually low masses of the bosons of the octet, since they are the eight Goldstone bosons corresponding to the spontaneously broken chiral symmetry group SU(3)$_{A}$ under which the quark fields transform; the second term acts on the $\eta $ and $\eta '$ and relates to the non-negligible $q\bar q$ annihilation effects \cite{DeRujula:1975ge} that arise when the lowest mesons are flavor diagonal.

They consider flavor states in the ideal mixing hypothesis, i.e. states with defined number of strange quarks and antiquarks, except for the lowest pseudoscalar nonet.
The ideal mixing is essentially a consequence of the OZI rule, introduced by Okubo \cite{Okubo:1963fa}, Zweig \cite{Zweig:1964jf} and Iizuka \cite{Iizuka:1966wu}. This hypothesis remains to be proved, but it is used by all the authors working on $q\bar q$ mesons and also on tetraquarks (see for example Jaffe \cite{Jaffe:1976ih,Jaffe:1976ig,Jaffe:2004ph} and Maiani et al. \cite{Maiani:2004uc}).

\begin{figure}
\caption{\label{fig:GraficoMesoni} The predicted mesonic masses (solid bars) confronted with the experimental data from the PDG \cite{PDG}, reported with their errors (gray boxes).}
\includegraphics[width=17cm]{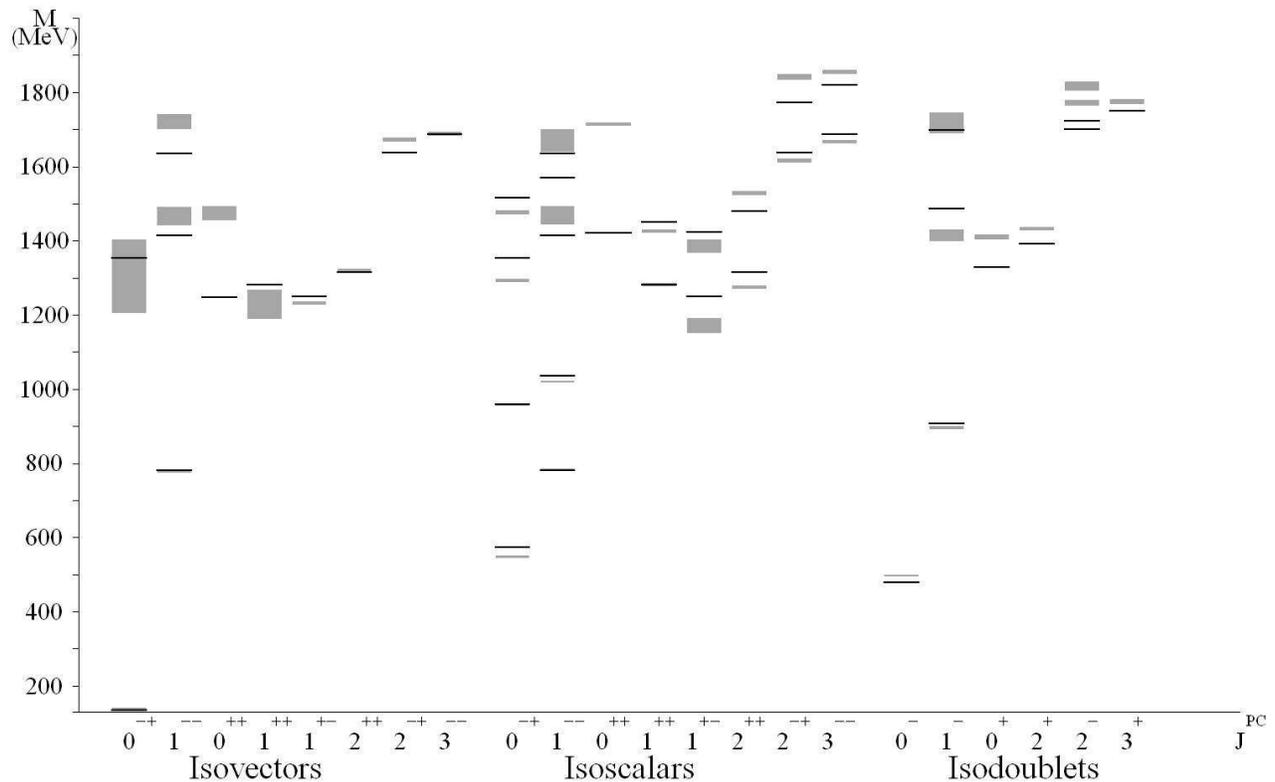}  
\end{figure}

Using the updated values for the light $q \bar q$ mesons reported by the last PDG \cite{PDG} (see Tables \ref{tab:mesonipi}, \ref{tab:mesonieta} and \ref{tab:mesonikappa}, see also Fig. \ref{fig:GraficoMesoni}) the results of the fit of the parameters, $M_{n}$, $M_{s}$, $a$, $b$, $c$, $d$, $h$, $i$, are 
\begin{subequations}
\begin{eqnarray}
 M_{n} & = & (0.3331\pm 0.0003)\;GeV  \\ 
 M_{s} & = & (0.4761\pm 0.0003)\;GeV  \\ 
 a & = & (1.389\pm 0.007)\;GeV^{2}  \\ 
 b & = & (1.0309\pm 0.0020)\;GeV^{2}  \\ 
 c & = &(0.079\pm 0.007)\;GeV^{2}  \\ 
 d & = &(0.0873\pm 0.0026)\;GeV^{2} \\  
 h & = &(0.4261\pm 0.0008)\;GeV^{2}  \\ 
 i & = &(0.1257\pm 0.0010)\;GeV^{2}.  
\end{eqnarray}
\end{subequations}

The data reported in the latest PDG are considerably different from those reported 15 years ago in PDG(1990) \cite{Hernandez:1990yc}. Moreover, some mesons that were not included in the original fit because they were 
poorly known at that time, now correspond to well measured resonances and have been included.

\begin{longtable}{|ccccccc|}
\caption{\label{tab:mesonipi}Experimental masses of the $\pi $ family mesons and predicted theoretical values.
Part (a) of the Table reports the mesons (and their squared masses) included in the fit and the squared masses predicted through Equation \ref{eq:formulamassa}. In part (b) there are some states, not included in the fit, for which we have predicted the masses. These states are written using the notation $\nu ^{2S+1}L_{J}$, where $\nu $ is a vibrational quantum number, $L$ is the relative orbital quantum number and $S$ the total spin. Candidate mesons seen experimentally have been assigned to the predicted states.} \\
\hline
\hline
Meson & $M^{2}$(exp.) ($GeV^{2}$)& $M^{2}$(teo.) ($GeV^{2}$)& $\nu $ & $L$ & $S$ & $J^{PC}$ \\
\hline
\hline
\endfirsthead
\caption[]{(continued)}\\
\hline
\hline
Meson & $M^{2}$(exp.) ($GeV^{2}$)& $M^{2}$(teo.) ($GeV^{2}$)& $\nu $ & $L$ & $S$ & $J^{PC}$ \\
\hline
\hline
\endhead
\hline
\multicolumn{7}{|c|}%
                   {(continued)} \\
\hline
\endfoot
\hline
\multicolumn{7}{|c|}%
                   {(end of table)} \\
\hline
\endlastfoot
\multicolumn{7}{|c|}{(a)} \\
\hline
$\pi $(140) & 0.01822521$\pm $0.00000002 & 0.018 & 0 & 0 & 0 & $0^{-+}$ \\
\hline
$\rho $(770) & 0.6019$\pm $0.0006 & 0.610 & 0 & 0 & 1 & $1^{--}$ \\
\hline
a$_{0}$(1450) & 2.17$\pm $0.08 & 1.554 & 0 & 1 & 1 & $0^{++}$ \\
\hline
a$_{1} $(1260) & 1.51$\pm $0.12 & 1.641 & 0 & 1 & 1 & $1^{++}$ \\
\hline
b$_{1} $(1235) & 1.517$\pm $0.010 & 1.562 & 0 & 1 & 0 & $1^{+-}$ \\
\hline
a$_{2} $(1320) & 1.737$\pm $0.002 & 1.728 & 0 & 1 & 1 & $2^{++}$ \\
\hline
$\rho $(1700) & 2.96$\pm $0.12 & 2.672 & 0 & 2 & 1 & $1^{--}$ \\
\hline
$\pi _{2} $(1670) & 2.797$\pm $0.018 & 2.680 & 0 & 2 & 0 & $2^{-+}$ \\
\hline
$\rho _{3}$(1690) & 2.852$\pm $0.012 & 2.847 & 0 & 2 & 1 & $3^{--}$ \\
\hline
$\pi  $(1300) & 1.7$\pm $0.3 & 1.833 & 1 & 0 & 0 & $0^{-+}$ \\
\hline
$\rho $(1450) & 2.15$\pm $0.11 & 1.999 & 1 & 0 & 1 & $1^{--}$ \\
\hline
 \multicolumn{7}{|c|}{(b)} \\
\hline
$0^{3}D_{2} $ & & 2.759 & 0 & 2 & 1 & $2^{--}$ \\
\hline
$0^{3}F_{2} $ & & 3.790 & 0 & 3 & 1 & $2^{++}$ \\
\hline
$0^{3}F_{3} $ & & 3.877 & 0 & 3 & 1 & $3^{++}$ \\
\hline
$0^{1}F_{3} $ & & 3.798 & 0 & 3 & 0 & $2^{+-}$ \\
\hline
$0^{3}F_{4} $ (a$_{4}$(2040))& 4.04$\pm$ 0.05 & 3.965 & 0 & 3 & 1 & $4^{++}$ \\
\hline
$1^{3}P_{0} $ & & 2.943 & 1 & 1 & 1 & $0^{++}$ \\
\hline
$1^{3}P_{1} $ (a$_{1}$(1640))& 2.71$\pm $0.07 & 3.030 & 1 & 1 & 1 & $1^{++}$ \\
\hline
$1^{1}P_{1} $ & & 2.951 & 1 & 1 & 0 & $1^{+-}$ \\
\hline
$1^{3}P_{2} $ (a$_{2}$(1700))& 3.00$\pm $0.06 & 3.117 & 1 & 1 & 1 & $2^{++}$ \\
\hline
$1^{3}D_{1} $ ($\rho $(2150))& 4.62$\pm $0.07 & 4.061 & 1 & 2 & 1 & $1^{--}$ \\
\hline
$1^{3}D_{2} $ & & 4.148 & 1 & 2 & 1 & $2^{--}$ \\
\hline
$1^{1}D_{2} $ ($\pi _{2}$(2100))& 4.41$\pm $0.12 & 4.069 & 1 & 2 & 0 & $2^{-+}$ \\
\hline
$1^{3}D_{3} $ ($\rho _{3}$(1990))& 3.93$\pm $0.06 & 4.236 & 1 & 2 & 1 & $3^{--}$ \\
\hline
$1^{3}F_{2} $ & & 5.179 & 1 & 3 & 1 & $2^{++}$ \\
\hline
$2^{3}S_{1} $ ($\rho $(1900))& $\sim $3.61 & 3.387 & 2 & 0 & 1 & $1^{--}$ \\
\hline
\end{longtable}

\begin{longtable}{|cccccccp{1.5cm}|}
\caption{\label{tab:mesonieta}Experimental masses of the $\eta $ family mesons and predicted  values. In the last column we report the mixing type: we adopt the ideal mixing hypothesis for all mesons, with the exception of the pseudoscalar mesons for which the mixing angle is indicated explicitly.} \\
\hline
\hline
Meson & $M^{2}$(exp.) ($GeV^{2}$)& $M^{2}$(teo.) ($GeV^{2}$)& $\nu $ & $L$ & $S$ & $J^{PC}$ & mixing type \\
\hline
\hline
\endfirsthead
\caption[]{(continued)}\\
\hline
\hline
Meson & $M^{2}$(exp.) ($GeV^{2}$)& $M^{2}$(teo.) ($GeV^{2}$)& $\nu $ & $L$ & $S$ & $J^{PC}$ & mixing type \\
\hline
\hline
\endhead
\hline
\multicolumn{8}{|c|}%
                   {(continued)} \\
\hline
\endfoot
\hline
\multicolumn{8}{|c|}%
                   {(end of table)} \\
\hline
\endlastfoot
 \multicolumn{8}{|c|}{(a)} \\
\hline
$\eta $(550) & 0.29954$\pm $0.00007 & 0.330 & 0 & 0 & 0 & $0^{-+}$ & $\theta _{fit}$=-17$^{\circ }$ \\
\hline
$\eta '$(958) & 0.9173$\pm $0.0002 & 0.809 & 0 & 0 & 0 & $0^{-+}$ & $\theta _{fit}$=-17$^{\circ }$ \\
\hline
$\omega  $(782) & 0.61242$\pm $0.00010 & 0.610 & 0 & 0 & 1 & $1^{--}$ & $n\bar n$ \\
\hline
$\phi  $(1020) & 1.03929$\pm $0.00004 & 1.073 & 0 & 0 & 1 & $1^{--}$ & $s\bar s$ \\
\hline
f$_{0} $(1710) & 2.94$\pm $0.03 & 2.017 & 0 & 1 & 1 & $0^{++}$ & $s\bar s$ \\
\hline
f$_{1} $(1285) & 1.643$\pm $0.002 & 1.641 & 0 & 1 & 1 & $1^{++}$ & $n\bar n$ \\
\hline
f$_{1} $(1420) & 2.033$\pm $0.004 & 2.104 & 0 & 1 & 1 & $1^{++}$ & $s\bar s$ \\
\hline
h$_{1} $(1170) & 1.37$\pm $0.05 & 1.562 & 0 & 1 & 0 & $1^{+-}$ & $n\bar n$ \\
\hline
h$_{1} $(1380) & 1.92$\pm $0.07 & 2.025 & 0 & 1 & 0 & $1^{+-}$ & $s\bar s$ \\
\hline
f$_{2} $(1270) & 1.626$\pm $0.004 & 1.728 & 0 & 1 & 1 & $2^{++}$ & $n\bar n$ \\
\hline
f'$_{2} $(1525) & 2.334$\pm $0.023 & 2.191 & 0 & 1 & 1 & $2^{++}$ & $s\bar s$ \\
\hline
$\omega _{2} $(1650) & 2.79$\pm $0.17 & 2.672 & 0 & 2 & 1 & $1^{--}$ & $n\bar n$ \\
\hline
$\eta _{2} $(1645) & 2.611$\pm $0.026 & 2.680 & 0 & 2 & 0 & $2^{-+}$ & $n\bar n$ \\
\hline
$\eta _{2} $(1870) & 3.39$\pm $0.05 & 3.143 & 0 & 2 & 0 & $2^{-+}$ & $s\bar s$ \\
\hline
$\omega _{3} $(1670) & 2.779$\pm $0.022 & 2.847 & 0 & 2 & 1 & $3^{--}$ & $n\bar n$ \\
\hline
$\phi _{3} $(1850) & 3.44$\pm $0.05 & 3.310 & 0 & 2 & 1 & $3^{--}$ & $s\bar s$ \\
\hline
$\eta $(1295) & 1.672$\pm $0.013 & 1.833 & 1 & 0 & 0 & $0^{-+}$ & $n\bar n$ \\
\hline
$\eta' $(1475) & 2.179$\pm $0.017 & 2.296 & 1 & 0 & 0 & $0^{-+}$ & $s\bar s$ \\
\hline
$\omega $(1420) & 2.01$\pm $0.09 & 1.999 & 1 & 0 & 1 & $1^{--}$ & $n\bar n$ \\
\hline
$\phi $(1680) & 2.82$\pm $0.11 & 2.462 & 1 & 0 & 1 & $1^{--}$ & $s\bar s$ \\
\hline
 \multicolumn{8}{|c|}{(b)} \\
\hline
$0^{3}P_{0} $ (f$_{0}$(1370))& 1.44-2.25 & 1.554 & 0 & 1 & 1 & $0^{++}$ & $n\bar n$ \\
\hline
$0^{3}D_{1} $ & & 3.135 & 0 & 2 & 1 & $1^{--}$ & $s\bar s$ \\
\hline
$0^{3}D_{2} $ & & 2.759 & 0 & 2 & 1 & $2^{--}$ & $n\bar n$ \\
\hline
$0^{3}D_{2} $ & & 3.222 & 0 & 2 & 1 & $2^{--}$ & $s\bar s$ \\
\hline
$0^{3}F_{2} $ (f$_{2}$(1910))& 3.667$\pm $0.027 & 3.790 & 0 & 3 & 1 & $2^{++}$ & $n\bar n$ \\
\hline
$0^{3}F_{2} $ (f$_{2}$(2150))& 4.65$\pm $0.10 & 4.253 & 0 & 3 & 1 & $2^{++}$ & $s\bar s$ \\
\hline
$0^{3}F_{3} $ & & 3.877 & 0 & 3 & 1 & $3^{++}$ & $n\bar n$ \\
\hline
$0^{3}F_{3} $ & & 4.340 & 0 & 3 & 1 & $3^{++}$ & $s\bar s$ \\
\hline
$0^{1}F_{3} $ & & 3.798 & 0 & 3 & 0 & $2^{+-}$ & $n\bar n$ \\
\hline
$0^{1}F_{3} $ & & 4.261 & 0 & 3 & 0 & $2^{+-}$ & $s\bar s$ \\
\hline
$0^{3}F_{4} $ (f$_{4}$(2050))& 4.14$\pm $0.04 & 3.965 & 0 & 3 & 1 & $4^{++}$ & $n\bar n$ \\
\hline
$0^{3}F_{4} $ & & 4.428 & 0 & 3 & 1 & $4^{++}$ & $s\bar s$ \\
\hline
$1^{3}S_{1} $ & & 1.999 & 1 & 0 & 1 & $1^{--}$ & $n\bar n$ \\
\hline
$1^{3}P_{0} $ & & 2.943 & 1 & 1 & 1 & $0^{++}$ & $n\bar n$ \\
\hline
$1^{3}P_{0} $ & & 3.406 & 1 & 1 & 1 & $0^{++}$ & $s\bar s$ \\
\hline
$1^{3}P_{1} $ & & 3.030 & 1 & 1 & 1 & $1^{++}$ & $n\bar n$ \\
\hline
$1^{3}P_{1} $ & & 3.493 & 1 & 1 & 1 & $1^{++}$ & $s\bar s$ \\
\hline
$1^{1}P_{1} $ & & 2.951 & 1 & 1 & 0 & $1^{+-}$ & $n\bar n$ \\
\hline
$1^{1}P_{1} $ & & 3.414 & 1 & 1 & 0 & $1^{+-}$ & $s\bar s$ \\
\hline
$1^{3}P_{2} $ (f$_{2}$(1640))& 2.683$\pm $0.020 & 3.117 & 1 & 1 & 1 & $2^{++}$ & $n\bar n$ \\
\hline
$1^{3}P_{2} $ (f$_{2}$(1950))& 3.78$\pm $0.05 & 3.580 & 1 & 1 & 1 & $2^{++}$ & $s\bar s$ \\
\hline
$1^{3}D_{1} $ & & 4.061 & 1 & 2 & 1 & $1^{--}$ & $n\bar n$ \\
\hline
$1^{3}D_{1} $ & & 4.524 & 1 & 2 & 1 & $1^{--}$ & $s\bar s$ \\
\hline
$1^{3}D_{2} $ & & 4.148 & 1 & 2 & 1 & $2^{--}$ & $n\bar n$ \\
\hline
$1^{3}D_{2} $ & & 4.611 & 1 & 2 & 1 & $2^{--}$ & $s\bar s$ \\
\hline
$1^{1}D_{2} $ & & 4.069 & 1 & 2 & 0 & $2^{-+}$ & $n\bar n$ \\
\hline
$1^{1}D_{2} $ & & 4.532 & 1 & 2 & 0 & $2^{-+}$ & $s\bar s$ \\
\hline
$1^{3}D_{3} $ & & 4.236 & 1 & 2 & 1 & $3^{--}$ & $n\bar n$ \\
\hline
$1^{3}D_{3} $ & & 4.698 & 1 & 2 & 1 & $3^{--}$ & $s\bar s$ \\
\hline
$1^{3}F_{2} $ (f$_{2}$(2300))& 5.28$\pm $0.13 & 5.179 & 1 & 3 & 1 & $2^{++}$ & $n\bar n$ \\
\hline
$1^{3}F_{2} $ (f$_{2}$(2340))& 5.47$\pm $0.28 & 5.642 & 1 & 3 & 1 & $2^{++}$ & $s\bar s$ \\
\hline
$2^{1}S_{0} $ ($\eta $(1760))& 3.10$\pm $0.04 & 3.222 & 2 & 0 & 0 & $0^{-+}$ & $n\bar n$ \\
\hline
$2^{3}P_{0} $ (f$_{0}$(2020))& 3.97$\pm $0.06 & 4.332 & 2 & 1 & 1 & $0^{++}$ & $n\bar n$ \\
\hline
$2^{3}P_{0} $ (f$_{0}$(2200))& 4.83$\pm $0.07 & 4.795 & 2 & 1 & 1 & $0^{++}$ & $s\bar s$ \\
\end{longtable}

\begin{table*}[!ht]
\begin{center}
\caption{\label{tab:mesonikappa}Experimental masses of the K family mesons compared with the theoretical values.}
\begin{tabular}{|ccccccc|}
\hline
Meson & $M^{2}$(exp.) ($GeV^{2}$)& $M^{2}$(teo.) ($GeV^{2}$)& $\nu $ & $L$ & $S$ & $J^{P}$ \\
\hline
 \multicolumn{7}{|c|}{(a)} \\
\hline
k(500) & 0.24768$\pm $0.00001 & 0.229 & 0 & 0 & 0 & $0^{-}$ \\
\hline
K$^{*}$(892) & 0.8032$\pm $0.0004 & 0.821 & 0 & 0 & 1 & $1^{-}$ \\
\hline
K$^{*}_{0}$(1430) & 1.99$\pm $0.02 & 1.765 & 0 & 1 & 1 & $0^{+}$ \\
\hline
K$^{*}_{2}$(1430) & 2.052$\pm $0.005 & 1.939 & 0 & 1 & 1 & $2^{+}$ \\
\hline
K$^{*}$(1680) & 2.95$\pm $0.16 & 2.883 & 0 & 2 & 1 & $1^{-}$ \\
\hline
K$_{2}$(1820) & 3.30$\pm $0.09 & 2.970 & 0 & 2 & 1 & $2^{-}$ \\
\hline
K$_{2}$(1770) & 3.14$\pm $0.05 & 2.891 & 0 & 2 & 0 & $2^{-}$ \\
\hline
K$^{*}_{3}$(1780) & 3.15$\pm $0.04 & 3.058 & 0 & 2 & 1 & $3^{-}$ \\
\hline
K$^{*}$(1410) & 2.00$\pm $0.06 & 2.210 & 1 & 0 & 1 & $1^{-}$ \\
\hline
 \multicolumn{7}{|c|}{(b)} \\
\hline
$0^{3}P_{1} $ &  & 1.852 & 0 & 1 & 1 & $1^{+}$ \\
\hline
$0^{1}P_{1} $ &  & 1.773 & 0 & 1 & 1 & $1^{+}$ \\
\hline
$0^{3}F_{2} $ & & 4.001 & 0 & 3 & 1 & $2^{+}$ \\
\hline
$0^{3}F_{3} $ & & 4.088 & 0 & 3 & 1 & $3^{+}$ \\
\hline
$0^{1}F_{3} $ & & 4.009 & 0 & 3 & 0 & $2^{+}$ \\
\hline
$0^{3}F_{4} $ (K$^{*}_{4}$(2045))& 4.18$\pm $0.04 & 4.176 & 0 & 3 & 1 & $4^{+}$ \\
\hline
$1^{1}S_{0} $ (K(1460))& $\sim $2.13 & 2.044 & 1 & 0 & 0 & $0^{-}$ \\
\hline
$1^{3}P_{0} $ & & 3.154 & 1 & 1 & 1 & $0^{+}$ \\
\hline
$1^{3}P_{1} $ & & 3.241 & 1 & 1 & 1 & $1^{+}$ \\
\hline
$1^{1}P_{1} $ (K$_{1}$(1650))& 2.72$\pm $0.17 & 3.162 & 1 & 1 & 0 & $1^{+}$ \\
\hline
$1^{3}P_{2} $ (K$^{*}_{2}$(1980))& 3.89$\pm $0.03 & 3.328 & 1 & 1 & 1 & $2^{+}$ \\
\hline
$1^{3}D_{1} $ & & 4.272 & 1 & 2 & 1 & $1^{-}$ \\
\hline
$1^{3}D_{2} $ & & 4.359 & 1 & 2 & 1 & $2^{-}$ \\
\hline
$1^{1}D_{2} $ & & 4.280 & 1 & 2 & 0 & $2^{-}$ \\
\hline
$1^{3}D_{3} $ & & 4.447 & 1 & 2 & 1 & $3^{-}$ \\
\hline
$1^{3}F_{2} $ & & 5.390 & 1 & 3 & 1 & $2^{+}$ \\
\hline
$1^{3}F_{3} $ (K$_{3}$(2320))& 5.40$\pm $0.11 & 5.477 & 1 & 3 & 1 & $3^{+}$ \\
\hline
\end{tabular}
\end{center}
\end{table*}

\clearpage

As expected, the mesons predicted by the Iachello mass formula reproduce the linear Regge trajectories, representations of $SO(4)$, the linearity of which is satisfied to a high accuracy for light mesons.
It is well known that the Regge behaviour \cite{Regge:1959mz} can be explained by means of string-like models \cite{Johnson:1975sg,'tHooft:1974hx}.

\section{\label{subsec:tetraformula} The $qq\bar q\bar q$ spectrum.}
A candidate tetraquark nonet was proposed in the 1970s by Jaffe \cite{Jaffe:1976ih,Jaffe:1976ig}. 
This nonet, with quantum numbers $J^{PC}=0^{++}$, includes the mesons $a_{0}(980)$, $f_{0}(980)$, $f_{0}(600)$ (also called $\sigma $ meson) and $\kappa (800)$.
We hypothesize, as did Jaffe \cite{Jaffe:1976ih,Jaffe:1976ig,Jaffe:2004ph}, Amsler and Tornqvist \cite{Amsler:2004ps}, Maiani \cite{Maiani:2004uc} and others, that this nonet is the fundamental tetraquark nonet, with total orbital angular momentum and total spin equal to zero.
The candidate tetraquark nonet quantum numbers are presented in Table \ref{tab:nonettotetra2}, where $N_{s}$ means the number of strange quarks and antiquarks.
\begin{table}
\begin{center}
\caption{\label{tab:nonettotetra2} Quantum numbers of the candidate tetraquark nonet}
\begin{tabular}{|c|c|c|c|c|}
\hline
Meson & Mass ($GeV$)& $N_{s}$ & $I^{G}(J^{PC})$ & Source \\
\hline
$a_{0}(980)$ & $0.9847\pm 0.0012$ & 2 & $1^{-}(0^{++})$ & PDG \cite{PDG} \\
\hline
$f_{0}(980)$ & $0.980\pm 0.010$ & 2 & $0^{+}(0^{++})$ &  PDG \cite{PDG}  \\
\hline
$f_{0}(600)$ & $0.478\pm 0.024$ & 0 & $0^{+}(0^{++})$ &  KLOE \cite{Aloisio:2002bt}  \\
\hline
$\kappa (800)$ & $0.797\pm 0.019$ & 1 & $\frac{1}{2}(0^{+})$ & E791 \cite{Aitala:2002kr}  \\
\hline
\end{tabular}
\end{center}
\end{table}

The Iachello, Mukhopadhyay and Zhang mass formula was originally developed for $q\bar q$ mesons. In order to describe uncorrelated tetraquark systems by means of an algebraic model  one should introduce a new spectrum generating algebra for the spatial part, in this case U(10), since we have nine spatial degrees of freedom. We will not address this difficult problem in this article, but we choose to write the part of the mass formula regarding the internal degrees of freedom in the same way. In $q\bar q$ mesons the splitting inside a given flavor multiplet to which is also associated a given spin, can be well described by means of the part of the Iachello, Mukhopadhyay and Zhang mass formula that depends only on the numbers of strange and non-strange quarks and antiquarks. 
It is not necessary, for the purpose of determining the mass splitting of the candidate tetraquark nonet, to calculate the spatial part of the mass formula; we can simply use 
 
\begin{equation}
\label{eq:tetrascorrelati}
M^{2}=\alpha +(N_{n}M_{n}+N_{s}M_{s})^{2},
\end{equation}
where $\alpha $ is a constant that encodes all the spatial and spin dependence of the mass formula, and $M_{n}$ and $M_{s}$ are the masses of the constituent quarks (as obtained from an upgrade of the fit of the Iachello, Mukhopadhyay and Zhang mass formula to the new PDG data \cite{PDG} on $q \bar q$ mesons).
We determine $\alpha $ by applying Equation (\ref{eq:tetrascorrelati}) to a well-known candidate tetraquark, $a_{0}(980)$, and in this way we set the energy scale. The value found is
\begin{equation}
\alpha =-1.650\;GeV^{2}
\end{equation}
With this value of $\alpha $ we predict the masses of the other mesons belonging to the same tetraquark nonet
\begin{equation}
M(\kappa (800))=0.726\;GeV
\end{equation}
\begin{equation}
M(f_{0}(600))=0.354\;GeV
\end{equation}
\begin{equation}
M(f_{0}(980))=0.984\;GeV
\end{equation}

\begin{figure}
\caption{Schematic graph of the fundamental tetraquark nonet.  The theoretical masses (in $MeV$), predicted according to Equation (\ref{eq:tetrascorrelati}), are reported below each resonance.}
\includegraphics[width=9.5cm]{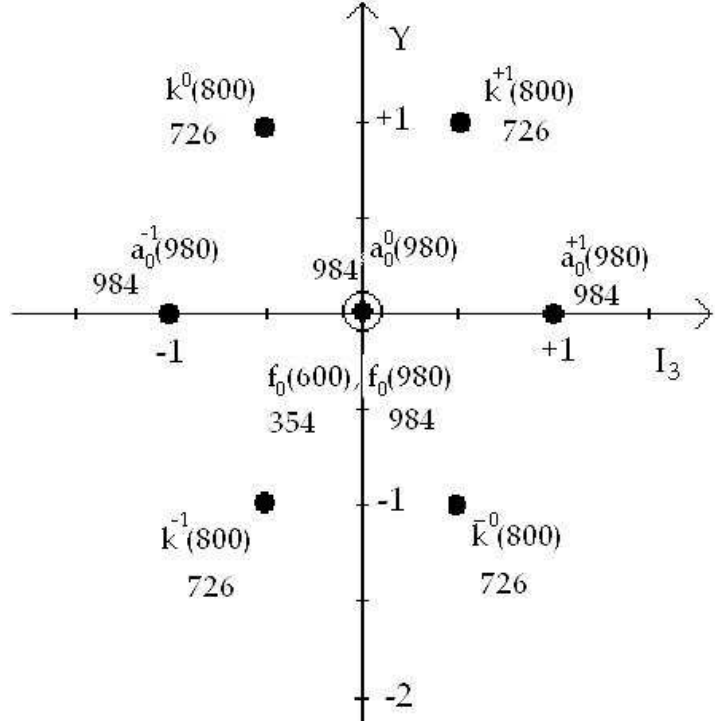}  
\end{figure}
The value of $f_{0}(980)$ agrees very well with the experimental mass reported by the PDG \cite{PDG}, on the contrary our masses of $f_{0}(600)$ and $\kappa (800)$ are respectively 5 and 4 experimental standard deviations from the values reported in Table \ref{tab:nonettotetra2}.
However, we must remember that the values of the masses $f_{0}(600)$ and $\kappa (800)$ found by the different experiments are scattered in a range of a few hundreds of $MeV$ around $500\;MeV$ and $800\;MeV$ respectively and the PDG does not report an average mass yet. Thus, new high statistics experiments for the $f_{0}(600)$ and $\kappa (800)$ are mandatory before reaching any conclusion.

\section{\label{sec:diqantidiq}Diquark-antidiquark model}

A diquark is a strongly correlated pair of quarks.
Since a pair of quarks cannot be a color singlet, the diquark can only be found confined into the hadrons and used as an effective degree of freedom. 

Recently many articles have been published regarding the open problem of diquark correlations both in baryons and tetraquarks. 
Different phenomenological indications for diquarks correlations have been collected over the years as pointed out in Ref. 
 \cite{Jaffe:2004ph} by Jaffe and in Ref. \cite{Selem:2006nd} by Selem and  Wilczek and references therein, moreover the occurence of rotational Regge trajectories for baryons with the same slope than the mesonic ones can be explained using a string model \cite{Johnson:1975sg} of the baryon, where at one end of the string there is a quark in $[3]_{c}$ color representation and at the other end a diquark in $[\bar 3]_{c}$.
Recently some papers, relating the physics of the instantons \cite{Schafer:2003nu,Faccioli:2004sn,Oka:2004qs,Schafer:1996wv}, and some calculations in Lattice QCD \cite{Alexandrou:2005zn,Walters:2002yh,Hess:1998sd} that support the existence of finite size diquarks as colour antitriplet bound states of two quarks have been published. The diquarks have also been studied in a Coulomb gauge QCD approach \cite{Alkofer:2005ug}, that proved their confinement and their well-defined size. One concern is that if diquark correlations are important for exotic states they would already be apparent in the ground state, positive parity nucleon. There is no clear evidence for diquarks in the nucleon, as stated in Refs. \cite{Glozman:1999pt,Leinweber:1993nr}, and surely completely not for pointlike ones.

From what we have written so far, it is clear that the existence of diquark correlations inside hadrons (and in particular ground state hadrons) is still an open problem. In the meantime, we believe it is not meaningless to study an effective diquark-antidiquark model for tetraquarks. Even if it will be finally found that quarks do not bind together, diquarks as effective degrees of freedom could be useful in hadron spectroscopy in order to correlate many data in terms of a phenomenological model. 

\subsection{\label{subsec:statidiqantiq}Classification of the tetraquark states in the diquark-antidiquark model.}
We think of the diquark as two correlated quark with no internal spatial excitations, or at least we hypothesize that their internal spatial excitations will be higher in energy than the scale of masses of the resonances we will consider.  We describe tetraquark mesons as being composed of a constituent diquark, $(qq)$, and a constituent antidiquark, $(\bar q \bar q)$.
The diquark SU(3)$_{c}$ color representations are $[\bar 3]_{c}$ and $[6]_{c}$, while the antidiquark ones are $[3]_{c}$ and $[\bar 6]_{c}$, using the standard convention of denoting  color and flavor by the dimensions of their representation. As the tetraquark must be a color singlet, the possible diquark-antidiquark color combinations are  
\begin{subequations} 
\begin{eqnarray}
&  \text{diquark\; in}\; [\bar 3]_{c},\; \text{antidiquark\; in}\; [3]_{c} & \\
&   \text{diquark\; in}\; [6]_{c},\; \text{antidiquark\; in}\; [\bar 6]_{c} &
\end{eqnarray}
\end{subequations}
Diquarks (and antidiquarks) are made up of two identical fermions and so they have to satisfy the Pauli principle. Since we consider diquarks with no internal spatial excitations, their color-spin-flavor wave functions must be antisymmetric. This limits the possible representations to being only 
\begin{subequations}
\begin{eqnarray}
&  \text{color \; in} ~ [\bar 3]~ \text{(AS),\; spin-flavor\;in} [21]_{sf}~\text{(S)} & \\
&  \text{color \;in }[6]~\text{(S)} \text{,\; spin-flavor\; in~} [15]_{sf}~ 
\text{(AS)}  & 
\end{eqnarray}
\end{subequations}
This is because we think of the diquark (antidiquark) as two correlated quarks (antiquarks) in an antisymmetric non-excited state. 
The decomposition of these SU$_{sf}$(6) representations in terms of SU(3)$_{f}\otimes $ SU(2)$_{s}$ is 
 (in the notation $[\text{flavor\;repr.,\;spin}]$)
\begin{subequations}
\begin{eqnarray}
& [21]_{sf}=[\bar 3,0]\oplus [6,1] & \\
& [15]_{sf}=[\bar 3,1]\oplus [6,0] &
\end{eqnarray} 
\end{subequations}
Using the notation $|\text{flavor\;repr.,\;color\;repr.,\;spin}\rangle$, the diquark states corresponding  to color 
$[\bar 3]_{c}$ and  $[6]_{c}$ respectively, are
\begin{eqnarray}
&  |[\bar 3]_{f},[\bar 3]_{c},0\rangle, |[6]_{f},[\bar 3]_{c},1\rangle & \\
&  |[\bar 3]_{f},[6]_{c},1\rangle, |[6]_{f},[6]_{c},0\rangle  & 
\end{eqnarray} 
The antidiquark states are obtained as the conjugate.  

In this paper we will consider only diquarks and antidiquarks in $[\bar 3]_{c}$ and $[3]_{c}$ color representations since like 
 Jaffe \cite{Jaffe:2004ph,Jaffe:1999ze} or Lichtenberg et al. \cite{Lichtenberg:1996fi}, for example, 
we expect that color-sextet diquarks will be higher in energy than color-triplet diquarks or even that they will not be bound at all. 

We have combined the allowed diquark and antidiquark states to derive the tetraquark color-spin flavor states; the situation is summarized in Table  \ref{tab:statiglobalidiq}. Since diquarks are considered with no internal spatial excitations, though this is an hypothesis in diquark-antidiquark models, their tetraquark states are a subset of the tetraquark states previously derived. In particular they corresponds to the subset with 
$L_{13}=L_{24}=0$,
where $L_{13}$ and $L_{24}$ are the relative orbital angular momenta of the two quarks and the two antiquarks respectively, and color 
$[\bar 3]_{c}\otimes [3]_{c}$. 
The relative orbital angular momentum among the diquark and the antidiquark is denoted by $L_{12-34}$; $S_{dq}$ and $ S_{d \bar q}$ are respectively the spin of the diquark and the spin of the antidiquark, and $S_{tot}$ is the total spin; $J$ is the total angular momentum.
 
\begin{table*}[!ht]
\begin{center}
\caption{\label{tab:statiglobalidiq}Diquark-antidiquark color, spin and flavor states. $S_{dq}$ ($S_{d\bar q}$) is the spin of the diquark (antidiquark) and $S_{tot}$ the total spin.}
\begin{tabular}{|c|c|c|c|c|c|}
\hline
color $(qq)\otimes (\bar q\bar q)$ & $S_{dq}$ & $S_{d\bar q}$ & flavor $(qq)\otimes (\bar q\bar q)$ & $S_{tot}$ & total flavor \\
\hline
$[\bar 3]_{c}\otimes [3]_{c}$ & 0 & 0 & $[\bar 3]_{f}\otimes [3]_{f}$ & 0 & $[1]\oplus [8]$ \\
\hline
$[\bar 3]_{c}\otimes [3]_{c}$ & 0 & 1 & $[\bar 3]_{f}\otimes [\bar 6]_{f}$ & 1 & $[8]\oplus [\overline{10}]$ \\
\hline
$[\bar 3]_{c}\otimes [3]_{c}$ & 1 & 0 & $[6]_{f}\otimes [3]_{f}$ & 1 & $[8]\oplus [10]$ \\
\hline
$[\bar 3]_{c}\otimes [3]_{c}$ & 1 & 1 & $[6]_{f}\otimes [\bar 6]_{f}$ & 0,1,2 & $[1]\oplus [8]\oplus [27]$  \\
\hline
\end{tabular}
\end{center}
\end{table*}
Table \ref{tab:statidiquark} shows the corresponding flavor tetraquark states for each diquark and antidiquark content in the ideal mixing hypothesis. 
Flavor exotic states (with $I>1$ and/or $|Y|>1$) are reported in \textbf{bold}. 
The notation used for diquarks should be explained. Scalar diquarks are represented by their constituent quarks
 (denoted by $s$ if strange, $n$ otherwise) in square brackets, while vector diquarks are in  curly brackets, since the explicit expression of diquarks is the commutator of the constituent quarks 
for the scalar ones and the anticommutator for the vector ones.

We can determine the $J^{PC}$ quantum numbers of the tetraquarks in the diquark-antidiquark limit 
starting from the possible quantum numbers classified for the uncorrelated tetraquark states
and applying the restrictions for the diquark-antidiquark limit, $L_{13}=L_{24}=0$ and color 
$[\bar 3]_{c}\otimes [3]_{c}$. With these restrictions the parity of a tetraquark in the diquark-antidiquark limit is
$P=(-1)^{L_{12-34}}$,
while the charge conjugation (obviously only for its eigenstates) is
$C=(-1)^{L_{12-34}+S_{tot}}$.
Consequently the $G$ parity is
$G=Ce^{i\pi I_{2}}=(-1)^{L_{12-34}+S_{tot}+I}$,
with the exceptions, already discussed in section \ref{subsec:paritàG}, of states belonging to $[8]\oplus [10]$ and $[8]\oplus [\overline{10}]$ flavor multiplets.
In Tables \ref{tab:spettroscopiadiquark1}, \ref{tab:spettroscopiadiquark2} and \ref{tab:spettroscopiadiquark3} we write the possible $J^{PC}$ combinations and diquark content of diquark-antidiquark systems with 
$L_{12-34}=0$, $L_{12-34}=1$ and $L_{12-34}=2$ respectively.
Exotic $J^{PC}$ combinations are in \textbf{bold}.

\begin{longtable}{|p{3.8cm}|c|p{2.3cm}|p{2.2cm}|p{2.2cm}|p{2.2cm}|p{1.5cm}|}
\caption{ \label{tab:statidiquark} Flavor diquark-antidiquark states in the \textquotedblleft ideal mixing\textquotedblright \, hypothesis (i.e. states with defined number of strange quarks plus antiquarks). In this table, for each different type of diquark and antidiquark, as reported in the first column with the notation $|\text{flavor\;repr.,\;color\;repr.,\;spin}\rangle$, one can read, starting from the third column, all the flavor states in the \textquotedblleft ideal mixing\textquotedblright \, hypothesis (ordered according to their different number $n_{s}$ of strange quarks plus antiquarks). For these states we use the notation $|I,I_{3},Y\rangle$ and under each its explicit diquark content (in terms of the flavor of the constituent quarks) is also reported. See also Appendix \ref{subapp:idealmixing} for an explicit expression of these states. Flavor exotic states are reported in \textbf{bold}}. \\
\hline
\hline
Diquark and & total & \multicolumn{5}{|c|}{flavor diquark-antidiquark states with defined $n_s$($n_{s}=0,1,2,3,4$)}  \\
antidiquark type& flavor& 0 & 1 & 2 & 3 & 4 \\
\hline
\hline
\endfirsthead
\caption[]{(continued)}\\
\hline
\hline
Diquark and & total & \multicolumn{5}{|c|}{flavor diquark-antidiquark states with defined $n_s$($n_{s}=0,1,2,3,4$)}  \\
antidiquark type& flavor& 0 & 1 & 2 & 3 & 4 \\
\hline
\hline
\endhead
\hline
\multicolumn{7}{|c|}%
                   {(continue)} \\
\hline
\endfoot
\hline
\multicolumn{7}{|c|}%
                   {(end of table)} \\
\hline
\endlastfoot
$|[\bar 3]_{f},[\bar 3]_{c},0\rangle|[3]_{f},[3]_{c},0\rangle$ & $[1]\oplus [8]$ & $|0,0,0\rangle\; \newline [n,n]\overline{[n,n]}$ & $|\frac{1}{2},I_{3},-1\rangle \; \newline
  [n,s]\overline{[n,n]}$ & $|0,0,0\rangle  \; \newline[n,s]\overline{[n,s]}$ &  &  \\

                                                   &                 &                 & $|\frac{1}{2},I_{3},+1\rangle ; \; \newline [n,n]\overline{[n,s]}$ & $|1,I_{3},0\rangle ; \;   \newline [n,s]\overline{[n,s]}$ &   &   \\
\hline
\hline
$|[\bar 3]_{f},[\bar 3]_{c},0\rangle|[\bar 6]_{f},[3]_{c},1\rangle$ & $[8]\oplus [\overline{10}]$ & $|1,I_{3},0\rangle\; \newline [n,n]\overline{\{ n,n \} }$ & $|\frac{1}{2},I_{3},-1\rangle;\; \newline [n,s]\overline{\{ n,n \} }$ & $|0,0,0\rangle;\; \newline [n,s]\overline{ \{ n,s \} }$ & $|\frac{1}{2},I_{3},+1\rangle\; \newline [n,s]\overline{ \{ s,s\} }$ &  \\

                                                        &                             &     & $|\frac{1}{2},I_{3},+1\rangle; \; \newline [n,n]\overline{\{ n,s \} }$ & $|1,I_{3},0\rangle; \; \newline [n,s]\overline{\{ n,s \} }$ &   &   \\

                                                        &                             &     & \bm{$|\frac{3}{2},I_{3},-1\rangle\;$} \newline \bm{$[n,s]\overline{\{ n,n \} }$} & \bm{$|0,0,+2\rangle\;$} \newline \bm{$[n,n]\overline{\{ s,s \} }$} &   &   \\
\hline
\hline
$|[6]_{f},[\bar 3]_{c},1\rangle|[3]_{f},[3]_{c},0\rangle$ & $[8]\oplus [10]$ & $|1,I_{3},0\rangle\; \newline \{ n,n \}\overline{[n,n] }$ & $|\frac{1}{2},I_{3},-1\rangle;\; \newline \{ n,s \}\overline{[n,n] }$ & $|0,0,0\rangle;\; \newline \{ n,s \}\overline{ [n,s] }$ & $|\frac{1}{2},I_{3},-1\rangle\; \newline \{ s,s\}\overline{ [n,s] }$ &  \\

                                                       &         &                         & $|\frac{1}{2},I_{3},+1\rangle;\; \newline\{ n,n \}\overline{[n,s] }$ & $|1,I_{3},0\rangle;\;\newline \{ n,s \}\overline{[n,s] }$ &   &   \\

                                                       &         &                         & \bm{$|\frac{3}{2},I_{3},+1\rangle\;$} \newline \bm{$ \{ n,n \}\overline{[n,s] }$} & \bm{$|0,0,-2\rangle\;$} \newline \bm{$ \{ s,s \}\overline{[n,n] }$} &   &   \\
\hline
\hline
$|[6]_{f},[\bar 3]_{c},1\rangle|[\bar 6]_{f},[3]_{c},1\rangle$ & $[1]\oplus [8]\oplus [27]$ & $|1,I_{3},0\rangle;\;\newline\{n,n\}\overline{\{ n,n \} }$ & $|\frac{1}{2},I_{3},-1\rangle;\;\newline\{ n,s\}\overline{\{ n,n \} }$ & $|0,0,0\rangle;\;\newline\{ n,s\} \overline{ \{ n,s \} }$      & $|\frac{1}{2},I_{3},-1\rangle;\;\newline\{ s,s\}\overline{ \{n,s\} }$ & $|0,0,0\rangle\;\newline\{ s,s\} \overline{ \{ s,s \} }$ \\

                                                   &                            & $|0,0,0\rangle;\;\newline\{n,n\}\overline{\{ n,n \} }$     & $|\frac{1}{2},I_{3},+1\rangle;\;\newline\{ n,n\}\overline{\{ n,s \} }$ & $|1,I_{3},0\rangle;\;\newline\{ n,s\} \overline{ \{ n,s \} }$  & $|\frac{1}{2},I_{3},+1\rangle\;\newline\{ n,s\}\overline{ \{s,s\} }$ &                                            \\

                                                   &                            & \bm{$|2,I_{3},0\rangle$} \newline \bm{$\{n,n\}\overline{\{ n,n \} }$} & \bm{$|\frac{3}{2},I_{3},+1\rangle$} \newline
\bm{$ \{ n,n\}\overline{\{ n,s \}} $} & \bm{$|1,I_{3},+2\rangle$} \newline \bm{$\{ n,n\} \overline{ \{ s,s \} }$} &                                                        &                                            \\

                                                   &                            &                                             & \bm{$|\frac{3}{2},I_{3},-1\rangle$} \newline \bf{$\{ n,s\}\overline{\{ n,n \} }$} & \bm{$|1,I_{3},-2\rangle$} \newline \bf{$\{ s,s\} \overline{ \{ n,n \} }$} &                                                        &                                            \\
\hline
\hline
\end{longtable}

\begin{table}[!ht]
\begin{center}
\caption{\label{tab:spettroscopiadiquark1} Spectroscopic classification and diquark content of tetraquarks states  with $L_{12-34}=0$ in the diquark-antidiquark limit.}
\begin{tabular}{|c|p{4.5cm}|}
\hline
$^{2S+1}L_{J}(J^{PC})$ & Diquark and antidiquark type \\
\hline
$^{1}S_{0}(0^{++})$ & $|[\bar 3]_{f},[\bar 3]_{c},0\rangle|[3]_{f},[3]_{c},0\rangle$; $|[6]_{f},[\bar 3]_{c},1\rangle|[\bar 6]_{f},[3]_{c},1\rangle$ \\
\hline
$^{3}S_{1}(1^{+})$ & $|[\bar 3]_{f},[\bar 3]_{c},0\rangle|[\bar 6]_{f},[3]_{c},1\rangle$; $|[6]_{f},[\bar 3]_{c},1\rangle|[3]_{f},[3]_{c},0\rangle$ \\
$^{3}S_{1}(1^{+-})$ & $|[6]_{f},[\bar 3]_{c},1\rangle|[\bar 6]_{f},[3]_{c},1\rangle$ \\
\hline
$^{5}S_{2}(2^{++})$ & $|[6]_{f},[\bar 3]_{c},1\rangle|[\bar 6]_{f},[3]_{c},1\rangle$ \\
\hline
\end{tabular}
\end{center}
\end{table}

\begin{table}[!ht]
\begin{center}
\caption{\label{tab:spettroscopiadiquark2} As in Table \ref{tab:spettroscopiadiquark1} but for $L_{12-34}=1$}
\begin{tabular}{|c|p{4.5cm}|}
\hline
$^{2S+1}L_{J}(J^{PC})$ & Diquark and antidiquark type \\
\hline
$^{1}P_{1}(1^{--})$ & $|[\bar 3]_{f},[\bar 3]_{c},0\rangle|[3]_{f},[3]_{c},0\rangle$; $|[6]_{f},[\bar 3]_{c},1\rangle|[\bar 6]_{f},[3]_{c},1\rangle$ \\
\hline
$^{3}P_{2}(2^{-})$  & $|[\bar 3]_{f},[\bar 3]_{c},0\rangle|[\bar 6]_{f},[3]_{c},1\rangle$; $|[6]_{f},[\bar 3]_{c},1\rangle|[3]_{f},[3]_{c},0\rangle$ \\
$^{3}P_{2}(2^{-+})$ & $|[6]_{f},[\bar 3]_{c},1\rangle|[\bar 6]_{f},[3]_{c},1\rangle$  \\
\hline
\bm{$^{3}P_{1}(1^{-})$}  & $|[\bar 3]_{f},[\bar 3]_{c},0\rangle|[\bar 6]_{f},[3]_{c},1\rangle$ ; $|[6]_{f},[\bar 3]_{c},1\rangle|[3]_{f},[3]_{c},0\rangle$ \\
\bm{$^{3}P_{1}(1^{-+})$} & $|[6]_{f},[\bar 3]_{c},1\rangle|[\bar 6]_{f},[3]_{c},1\rangle$  \\
\hline
$^{3}P_{0}(0^{-})$  & $|[\bar 3]_{f},[\bar 3]_{c},0\rangle|[\bar 6]_{f},[3]_{c},1\rangle$; $|[6]_{f},[\bar 3]_{c},1\rangle|[3]_{f},[3]_{c},0\rangle$ \\
$^{3}P_{0}(0^{-+})$ & $|[6]_{f},[\bar 3]_{c},1\rangle|[\bar 6]_{f},[3]_{c},1\rangle$  \\
\hline
$^{5}P_{3}(3^{--})$ & $|[6]_{f},[\bar 3]_{c},1\rangle|[\bar 6]_{f},[3]_{c},1\rangle$ \\
\hline
$^{5}P_{2}(2^{--})$ & $|[6]_{f},[\bar 3]_{c},1\rangle|[\bar 6]_{f},[3]_{c},1\rangle$ \\
\hline
$^{5}P_{1}(1^{--})$ & $|[6]_{f},[\bar 3]_{c},1\rangle|[\bar 6]_{f},[3]_{c},1\rangle$ \\
\hline
\end{tabular}
\end{center}
\end{table}

\begin{table}[!ht]
\begin{center}
\caption{\label{tab:spettroscopiadiquark3} As in Table \ref{tab:spettroscopiadiquark1} but for $L_{12-34}=2$}
\begin{tabular}{|c|p{4.5cm}|}
\hline
$^{2S+1}L_{J}(J^{PC})$ & Diquark and antidiquark type \\
\hline
$^{1}D_{2}(2^{++})$ & $|[\bar 3]_{f},[\bar 3]_{c},0\rangle|[3]_{f},[3]_{c},0\rangle$; $|[6]_{f},[\bar 3]_{c},1\rangle|[\bar 6]_{f},[3]_{c},1\rangle$ \\
\hline
$^{3}D_{3}(3^{+})$  & $|[\bar 3]_{f},[\bar 3]_{c},0\rangle|[\bar 6]_{f},[3]_{c},1\rangle$; $|[6]_{f},[\bar 3]_{c},1\rangle|[3]_{f},[3]_{c},0\rangle$ \\
$^{3}D_{3}(3^{+-})$ & $|[6]_{f},[\bar 3]_{c},1\rangle|[\bar 6]_{f},[3]_{c},1\rangle$ \\
\hline
\bm{$^{3}D_{2}(2^{+})$}  & $|[\bar 3]_{f},[\bar 3]_{c},0\rangle|[\bar 6]_{f},[3]_{c},1\rangle$; $|[6]_{f},[\bar 3]_{c},1\rangle|[3]_{f},[3]_{c},0\rangle$ \\
\bm{$^{3}D_{2}(2^{+-})$} & $|[6]_{f},[\bar 3]_{c},1\rangle|[\bar 6]_{f},[3]_{c},1\rangle$ \\
\hline
$^{3}D_{1}(1^{+})$  & $|[\bar 3]_{f},[\bar 3]_{c},0\rangle|[\bar 6]_{f},[3]_{c},1\rangle$; $|[6]_{f},[\bar 3]_{c},1\rangle|[3]_{f},[3]_{c},0\rangle$ \\
$^{3}D_{1}(1^{+-})$ & $|[6]_{f},[\bar 3]_{c},1\rangle|[\bar 6]_{f},[3]_{c},1\rangle$ \\
\hline
$^{5}D_{4}(4^{++})$ & $|[6]_{f},[\bar 3]_{c},1\rangle|[\bar 6]_{f},[3]_{c},1\rangle$ \\
\hline
$^{5}D_{3}(3^{++})$ & $|[6]_{f},[\bar 3]_{c},1\rangle|[\bar 6]_{f},[3]_{c},1\rangle$ \\
\hline
$^{5}D_{2}(2^{++})$ & $|[6]_{f},[\bar 3]_{c},1\rangle|[\bar 6]_{f},[3]_{c},1\rangle$ \\
\hline
$^{5}D_{1}(1^{++})$ & $|[6]_{f},[\bar 3]_{c},1\rangle|[\bar 6]_{f},[3]_{c},1\rangle$ \\
\hline
$^{5}D_{0}(0^{++})$ & $|[6]_{f},[\bar 3]_{c},1\rangle|[\bar 6]_{f},[3]_{c},1\rangle$ \\
\hline
\end{tabular}

\end{center}
\end{table}
How to read Tables \ref{tab:statidiquark}, \ref{tab:spettroscopiadiquark1}, \ref{tab:spettroscopiadiquark2}, \ref{tab:spettroscopiadiquark3} can be explained by examples.
In Table \ref{tab:statidiquark} we can read the diquark content of the states belonging to a given flavor multiplet. For example, as we can read from the first line of Table \ref{tab:statidiquark}, the nine states belonging to the flavor nonet are made up of two scalar diquarks.
In particular the state $|0,0,0\rangle$ contains the $[n,n]$ diquark and the $\overline{[n,n]}$ antidiquark. Tables \ref{tab:spettroscopiadiquark1}, \ref{tab:spettroscopiadiquark2}, \ref{tab:spettroscopiadiquark3} show for a given $J^{PC}$  which diquark-antidiquark type content is possible and also which  $J^{PC}$ quantum numbers can be assigned to a given diquark-antidiquark state.
For example, as indicated in the first line of Table \ref{tab:spettroscopiadiquark1}, the only possible tetraquarks with $^{1}S_{0}(0^{++})$ quantum numbers are those containing two scalar diquarks or two vector diquarks (which correspond respectively to the flavor nonet and the flavor 36-plet, as we can see in Table \ref{tab:statidiquark}). 

\subsection{\label{subsec:spettrodiqantiq}The tetraquark nonet spectrum in the diquark-antidiquark model.}
We describe diquark-antidiquark tetraquark configurations by using $U(4)\otimes SU(3)_{f}\otimes SU(2)_{s}\otimes SU(3)_{c}$ as spectrum generating algebra, by analogy with what was done by Iachello et al. \cite{Iachello:1991re,Iachello:1991fj} for the normal mesons. In a string model \cite{Johnson:1975sg,'tHooft:1974hx} the slopes of these trajectories depend only on the color representation of the constituent particles.
Thus the slope of Regge trajectories of tetraquarks made up of a diquark in $[\bar 3]_{c}$ and an antidiquark in $[3]_{c}$ is the same as the slope of Regge trajectories of $q\bar q$ mesons.

For the tetraquark in the diquark-antidiquark model, we can use the mass formula developed by Iachello et al. \cite{Iachello:1991re,Iachello:1991fj} for the normal mesons, but replacing the masses of the quark and the antiquark with those of the diquark and the antidiquark:
\begin{equation}
\label{eq:formulamassadqantidq}
M^{2}=(M_{qq}+M_{\bar q\bar q})^{2}+a\cdot n+b\cdot L_{12-34}+c\cdot S_{tot}+d\cdot J,
\end{equation}
where $M_{qq}$ and $M_{\bar q\bar q}$ are the diquark and antidiquark masses, $n$ is a vibrational quantum number, $L_{12-34}$ the relative orbital angular momentum, $S_{tot}$ the total spin and $J$ the total angular momentum.

We need, then, to determine the diquark masses. This can be done by fitting the mass formula (\ref{eq:formulamassadqantidq}) with the mass values of the tetraquark candidate nonet\footnote{This nonet has quantum numbers $n=L_{12-34}=S_{tot}=J=0$, so we do not need to know the parameters $a$, $b$, $c$ and $d$} $a_{0}(980)$, $f_{0}(980)$, $f_{0}(600)$ and $\kappa (800)$. 
Following Jaffe's arguments  \cite{Jaffe:1976ig,Jaffe:1999ze,Jaffe:2004ph}, the candidate tetraquark nonet is the fundamental tetraquark multiplet and it contains the lighter diquarks, i.e. scalar diquarks. 
\begin{table}
\begin{center}
\caption{\label{tab:nonettotetra2diq} Quantum numbers of the candidate tetraquark nonet. $\kappa (800)$ corresponds to $[n,n]\overline{[n,s]}$ and also to its conjugate.}
\begin{tabular}{|c|c|c|c|c|}
\hline
Meson & Mass ($GeV$)& Diquark content & $I^{G}(J^{PC})$ & Source \\
\hline
$a_{0}(980)$ & $0.9847\pm 0.0012$ & $[n,s]\overline{[n,s]}$ & $1^{-}(0^{++})$ & PDG \cite{PDG} \\
\hline
$f_{0}(980)$ & $0.980\pm 0.010$ & $[n,s]\overline{[n,s]}$ & $0^{+}(0^{++})$ &  PDG \cite{PDG}  \\
\hline
$f_{0}(600)$ & $0.478\pm 0.024$ & $[n,n]\overline{[n,n]}$ & $0^{+}(0^{++})$ &  KLOE \cite{Aloisio:2002bt}  \\
\hline
$\kappa (800)$ & $0.797\pm 0.019$ & $[n,n]\overline{[n,s]}$ & $\frac{1}{2}(0^{+})$ & E791 \cite{Aitala:2002kr}  \\
\hline
\end{tabular}
\end{center}
\end{table}

The masses of the scalar diquarks resulting from the fit are:
\begin{equation}
\label{eq:massadiquarkscalari}
M_{[n,n]}  =  0.275\;GeV,\;\;M_{[n,s]}  =  0.492\;GeV  
\end{equation}
The masses of the candidate tetraquark nonet obtained from the fit\footnote{This fit has been accomplished in a weighted way: for $a_{0}(980)$ and $f_{0}(980)$ the standard weights, coming from the inverse of the squared errors, have been used; the PDG does not give an estimate of the average values of the masses for $f_{0}(600)$ and $\kappa (800)$ and the values from different experiments are scattered in a range of $400\div 700\;MeV$ and $700\div 900\;MeV$ respectively, which have been used for the calculation of the weights, as an estimate of their unreliability.} are:
\begin{subequations}
\begin{eqnarray}
 & M_{a_{0}(980)}=M_{f_{0}(980)}= 0.984\;GeV & \\
 & M_{f_{0}(600)}= 0.550\;GeV & \\
 & M_{\kappa (800)}= 0.767\;GeV. & 
\end{eqnarray}
\end{subequations}
The masses of $a_{0}(980)$ and $f_{0}(980)$ agree with the experimental values reported by the PDG \cite {PDG}, $\kappa (800)$ and $f_{0}(600)$ are respectively within 2 and 3 experimental standard deviations from the values reported in Table \ref{tab:nonettotetra2diq}.

The value of $f_{0}(600)$ is similar to the value recently found by Mathur, Nilmani and others \cite{Mathur:2006bs} in a tetraquark model with Lattice QCD.  

\section{\label{sec:conclusione}Summary, conclusions and outlook}

In this work, we have constructed a complete classification scheme of the tetraquark states in terms of SU(6)$_{sf}$
spin-flavor multiplets, and their flavor and spin content in terms of SU(3)$_{f}$ and SU(2)$_{s}$ states. 
Moreover, we have discussed the permutation symmetry properties of both the spin-flavor and orbital parts of the
$qq$ and $\bar q\bar q$ subsystems. In order to obtain the total wave function, the spin-flavor part has been combined
with the color and orbital contributions in such a way that the total tetraquark wave function is a
color singlet and is antisymmetric under the interchange of the two quarks and the two antiquarks. This classification
scheme is general and complete, and may be helpful for experimental, CQM and lattice QCD
studies. In particular, the constructed basis for tetraquark states will enable the eigenvalue
problem to be solved for a definite dynamical model. 

As an application, we have calculated the mass spectrum of the candidate tetraquark nonet, adapting to the tetraquark case the Iachello, Mukhopadhyay and Zhang \cite{Iachello:1991re,Iachello:1991fj} mass formula, developed originally for the $q\bar q$ mesons. We have considered only the part of this formula that gives the splitting inside a given multiplet.

The predicted tetraquark states in the low energy range are much more numerous than the experimental candidate tetraquarks. 
So, if the tetraquark model is correct, we must solve the problem of the missing tetraquark resonances.

The introduction of the diquark-antidiquark model, in section \ref{sec:diqantidiq}, helps us to drastically reduce the number of predicted tetraquark states. In fact the allowed states in this model are only a small subset of the states in the uncorrelated model. Nevertheless this cut is not sufficient and the remaining tetraquark resonances are still too numerous.
Thus, we need another explanation for the missing resonances.

If it is heavy enough, a $qq\bar q\bar q$ meson will be unstable against decay into two $q\bar q$ mesons. 
The $qq\bar q\bar q$ state simply falls apart, or dissociates \cite{Jaffe:1976ih,Jaffe:1976ig}. Thus, we can deduce that if a given $qq\bar q\bar q$ state is above threshold for decay into a 
\textquotedblleft dissociation\textquotedblright \;channel, it is very broad into that channel and the higher the energy of the resonance is the broader the phase space is.

The great width of most $qq\bar q\bar q$ mesons will account for their experimental elusiveness, thus making it difficult to establish their resonant character at all. The $f_{0}(600)$ provides a clear example of this. 
Many higher-mass $qq\bar q\bar q$ states not only may be as broad and confusing as the $f_{0}(600)$, but also will probably occur in channels with substantial inelastic background obscuring their resonant behaviour. 

As an alternative to the tetraquark hypothesis, the possibility was considered that
$a_{0}(980)$ and $f_{0}(980)$ may be $K-\bar K$ bound states, kept together
by hadron exchange forces, the same that bind nucleons
in the nuclei, color singlet remnants of the confining
color forces (hence the name $K-\bar K$ molecules \cite{DeRujula:1976qd} used
in this connection). If they are indeed $K-\bar K$ molecules,
scalar mesons do not need to make a complete SU(3)$_{f}$ multiplet so that
this idea would be consistent with the lack
of evidence of $f_{0}(600)$ and $\kappa (800)$. However, if the existence of these particle were confirmed, it would be very hard to consider either of them as a $\pi -\pi $ or $\pi -K$ molecule, since the
latter particles would in any case lie considerably higher
than the respective thresholds.
We see that the existence or absence of these light scalars
is crucial in assessing the nature of $a_{0}(980)$ and $f_{0}(980)$.
From this point of view, the recent observations of
$f_{0}(600)$ and $\kappa (800)$ in D non-leptonic decays at Fermilab\cite{Aitala:2002kr} and in the $\pi \pi $ spectrum in $\phi \rightarrow \pi ^{0}\pi ^{0}\gamma $ at Frascati
\cite{Aloisio:2002bs,Aitala:2002kr} have considerably reinforced the hypothesis of a full tetraquark
nonet with inverted spectrum. The experimental situation and the latest results concerning $f_{0}(600)$ and $\kappa (800)$ are summarized in Refs. \cite{Caprini:2005zr,Bugg:2005xz,Ablikim:2005ni} and references therein.
New high-statistics experiments dedicated to these resonances are important in order to confirm or refute their existence. 

\begin{appendix}
\appendix
\section{\label{app:statisapore}The $qq\bar q\bar q$ flavor states}

In this appendix we write the $qq\bar q\bar q$ flavor states explicitly in terms of the states of the single quarks and antiquarks (the color  states may be easily obtained from the flavor ones by using the replacements $u\leftrightarrow r$, $d\leftrightarrow g$ and $s\leftrightarrow b$).
These states are calculated by using the SU(3) Clebsch-Gordan coefficients \cite{Kaeding,DeSwart} in the De Swart \cite{DeSwart} phase convention.

A generic flavor state is expressed by $|[R]I,I_{3},Y>$, where $[R]$ indicates the SU(3)$_{f}$ representation, $I$ the isospin quantum number, $I_{3}$ its third component and $Y$ the hypercharge.
The single quark states are written in short as $u$, $d$ and $s$, where
\begin{displaymath} 
u=|[3]\frac{1}{2},-\frac{1}{2},+\frac{1}{3}>
\end{displaymath}
\begin{displaymath} 
d=|[3]\frac{1}{2},+\frac{1}{2},+\frac{1}{3}>
\end{displaymath}
\begin{displaymath} 
s=|[3]0,0,-\frac{2}{3}>.
\end{displaymath}  

In this appendix the tetraquark states are written in the $(qq)(\bar q\bar q)$ configuration; thus, in addition to the representation $[R]$ to which the state belongs, we also show the representations $[R']$ and $[R'']$ respectively of the two quarks and the two antiquarks.
In short, a tetraquark state will be written as $|[R]I,I_{3},Y>_{R'\;R''}$. The states in the $q\bar qq\bar q$ configuration are a linear combination of those in the $(qq)(\bar q\bar q)$ configuration.

\subsection{\label{subapp:statisapore}The tetraquark flavor states in the $(qq)(\bar q\bar q)$ configuration}

\begin{eqnarray}
 & |[1]0,0,0>_{\bar 3\,3}  =  \frac{1}{2\sqrt{3}}(su\bar s\bar u-su\bar u\bar s-us\bar s\bar u+us\bar u\bar s-ds\bar d\bar s+  & \\  \nonumber
 & +ds\bar s\bar d+sd\bar d\bar s-sd\bar s\bar d+ud\bar u\bar d-ud\bar d\bar u-du\bar u\bar d+du\bar d\bar u) &
\end{eqnarray}
\begin{eqnarray}
 & |[1]0,0,0>_{6\,\bar 6}  =  \frac{1}{2\sqrt{6}}(2uu\bar u\bar u-ud\bar u\bar d-ud\bar d\bar u-du\bar u\bar d-du\bar d\bar u+  & \\  \nonumber
 & +2dd\bar d\bar d-us\bar u\bar s-us\bar s\bar u-su\bar u\bar s-su\bar s\bar u+ds\bar d\bar s+ds\bar s\bar d+sd\bar d\bar s+sd\bar s\bar d+2ss\bar s\bar s) &
\end{eqnarray}
\begin{equation}
|[8]1,+1,0>_{\bar 3\,3}=\frac{1}{2}(-su\bar d\bar s+su\bar s\bar d+us\bar d\bar s-us\bar s\bar d)
\end{equation}
\begin{eqnarray}
& |[8]1,0,0>_{\bar 3\,3}=\frac{1}{2\sqrt{2}}(-sd\bar d\bar s-su\bar u\bar s+sd\bar s\bar d+ & \\ \nonumber
& +su\bar s\bar u+ds\bar d\bar s+us\bar u\bar s-ds\bar s\bar d-us\bar s\bar u) &
\end{eqnarray}
\begin{equation}
|[8]1,-1,0>_{\bar 3\,3}=\frac{1}{2}(-ds\bar s\bar u+ds\bar u\bar s+sd\bar s\bar u-sd\bar u\bar s)
\end{equation} 
\begin{equation}
|[8]\frac{1}{2},+\frac{1}{2},+1>_{\bar 3\,3}=\frac{1}{2}(-ud\bar d\bar s+ud\bar s\bar d+du\bar d\bar s-du\bar s\bar d)
\end{equation}
\begin{equation}
|[8]\frac{1}{2},-\frac{1}{2},+1>_{\bar 3\,3}=\frac{1}{2}(-ud\bar u\bar s+ud\bar s\bar u+du\bar u\bar s-du\bar s\bar u)
\end{equation}
\begin{equation}
|[8]\frac{1}{2},+\frac{1}{2},-1>_{\bar 3\,3}=\frac{1}{2}(su\bar u\bar d-su\bar d\bar u-us\bar u\bar d+us\bar d\bar u)
\end{equation}
\begin{equation}
|[8]\frac{1}{2},-\frac{1}{2},-1>_{\bar 3\,3}=\frac{1}{2}(-ds\bar u\bar d+ds\bar d\bar u+sd\bar u\bar d-sd\bar d\bar u)
\end{equation}
\begin{eqnarray}
 & |[8]0,0,0>_{\bar 3\,3}  =  \frac{1}{2\sqrt{6}}(-su\bar s\bar u+su\bar u\bar s+us\bar s\bar u-us\bar u\bar s+ds\bar d\bar s+  & \\  \nonumber
 & -ds\bar s\bar d-sd\bar d\bar s+sd\bar s\bar d+2ud\bar u\bar d-2ud\bar d\bar u-2du\bar u\bar d+2du\bar d\bar u) &
\end{eqnarray}
\begin{eqnarray}
 & |[8]1,+1,0>_{6\,3} = \frac{1}{2\sqrt{3}}(-2uu\bar u\bar d+2uu\bar d\bar u-us\bar d\bar s+ & \\ \nonumber
      &  +us\bar s\bar d-su\bar d\bar s+su\bar s\bar d) & 
\end{eqnarray}
\begin{eqnarray}
 & |[8]1,0,0>_{6\,3} = \frac{1}{2\sqrt{6}}(-2du\bar u\bar d-2ud\bar u\bar d+2du\bar d\bar u+2ud\bar d\bar u+ & \\ \nonumber
      &  +ds\bar s\bar d+us\bar s\bar u-ds\bar d\bar s-us\bar u\bar s-sd\bar d\bar s-su\bar u\bar s+sd\bar s\bar d+su\bar s\bar u) & 
\end{eqnarray}
\begin{eqnarray}
& |[8]1,-1,0>_{6\,3} = \frac{1}{2\sqrt{3}}(2dd\bar d\bar u-2dd\bar u\bar d-ds\bar u\bar s+ & \\ \nonumber
      &  +ds\bar s\bar u-sd\bar u\bar s+sd\bar s\bar u) & 
\end{eqnarray}
\begin{eqnarray}
 & |[8]\frac{1}{2},+\frac{1}{2},+1>_{6\,3} = \frac{1}{2\sqrt{3}}(2uu\bar s\bar u-2uu\bar u\bar s+ud\bar d\bar s-ud\bar s\bar d+ & \\ \nonumber
      & +du\bar d\bar s-du\bar s\bar d ) & 
\end{eqnarray}
\begin{eqnarray}
 & |[8]\frac{1}{2},-\frac{1}{2},+1>_{6\,3} = \frac{1}{2\sqrt{3}}(2dd\bar d\bar s-2dd\bar s\bar d+du\bar s\bar u+ud\bar s\bar u+ & \\ \nonumber
      & -du\bar u\bar s-ud\bar u\bar s ) & 
\end{eqnarray}
\begin{eqnarray}
 & |[8]\frac{1}{2},+\frac{1}{2},-1>_{6\,3} = \frac{1}{2\sqrt{3}}(2ss\bar s\bar d-2ss\bar d\bar s+su\bar d\bar u-su\bar u\bar d+ & \\ \nonumber
      & +us\bar d\bar u-us\bar u\bar d ) & 
\end{eqnarray}
\begin{eqnarray}
 & |[8]\frac{1}{2},-\frac{1}{2},-1>_{6\,3} = \frac{1}{2\sqrt{3}}(2ss\bar s\bar u-2ss\bar u\bar s+sd\bar d\bar u-sd\bar u\bar d+ & \\ \nonumber
      & +ds\bar d\bar u-ds\bar u\bar d ) & 
\end{eqnarray}
\begin{eqnarray}
 & |[8]0,0,0>_{6\,3}  =  \frac{1}{2\sqrt{2}}(su\bar s\bar u-su\bar u\bar s+us\bar s\bar u-us\bar u\bar s+ds\bar d\bar s+  & \\  \nonumber
 & -ds\bar s\bar d+sd\bar d\bar s-sd\bar s\bar d) &
\end{eqnarray}
\begin{eqnarray}
 & |[8]1,+1,0>_{\bar 3\,\bar 6} = \frac{1}{2\sqrt{3}}(2ud\bar d\bar d-2du\bar d\bar d+us\bar d\bar s+ & \\ \nonumber
      &  +us\bar s\bar d-su\bar d\bar s-su\bar s\bar d) & 
\end{eqnarray}
\begin{eqnarray}
 & |[8]1,0,0>_{\bar 3\,\bar 6} = \frac{1}{2\sqrt{6}}(-2du\bar u\bar d+2ud\bar u\bar d-2du\bar d\bar u+2ud\bar d\bar u+ & \\ \nonumber
      &  +ds\bar s\bar d+us\bar s\bar u+ds\bar d\bar s+us\bar u\bar s-sd\bar d\bar s-su\bar u\bar s-sd\bar s\bar d-su\bar s\bar u) & 
\end{eqnarray}
\begin{eqnarray}
& |[8]1,-1,0>_{\bar 3\,\bar 6} = \frac{1}{2\sqrt{3}}(2ud\bar u\bar u-2du\bar u\bar u+ds\bar u\bar s+ & \\ \nonumber
      &  +ds\bar s\bar u-sd\bar u\bar s-sd\bar s\bar u) & 
\end{eqnarray}
\begin{eqnarray}
 & |[8]\frac{1}{2},+\frac{1}{2},+1>_{\bar 3\,\bar 6} = \frac{1}{2\sqrt{3}}(2us\bar s\bar s-2su\bar s\bar s+ud\bar d\bar s+ud\bar s\bar d+ & \\ \nonumber
      & -du\bar d\bar s-du\bar s\bar d ) & 
\end{eqnarray}
\begin{eqnarray}
 & |[8]\frac{1}{2},-\frac{1}{2},+1>_{\bar 3\,\bar 6} = \frac{1}{2\sqrt{3}}(2ds\bar s\bar s-2sd\bar s\bar s+ud\bar u\bar s-du\bar u\bar s+ & \\ \nonumber
      & +ud\bar s\bar u-du\bar s\bar u ) & 
\end{eqnarray}
\begin{eqnarray}
 & |[8]\frac{1}{2},+\frac{1}{2},-1>_{\bar 3\,\bar 6} = \frac{1}{2\sqrt{3}}(2sd\bar d\bar d-2ds\bar d\bar d-su\bar u\bar d+us\bar u\bar d+ & \\ \nonumber
      & -su\bar d\bar u-us\bar d\bar u ) & 
\end{eqnarray}
\begin{eqnarray}
 & |[8]\frac{1}{2},-\frac{1}{2},-1>_{\bar 3\,\bar 6} = \frac{1}{2\sqrt{3}}(2us\bar u\bar u-2su\bar u\bar u+sd\bar u\bar d+sd\bar d\bar u+ & \\ \nonumber
      & -ds\bar u\bar d-ds\bar d\bar u ) & 
\end{eqnarray}
\begin{eqnarray}
 & |[8]0,0,0>_{\bar 3\,\bar 6}  =  \frac{1}{2\sqrt{2}}(-su\bar s\bar u-su\bar u\bar s+us\bar s\bar u+us\bar u\bar s-ds\bar d\bar s+  & \\  \nonumber
 & -ds\bar s\bar d+sd\bar d\bar s+sd\bar s\bar d) &
\end{eqnarray}
\begin{eqnarray}
 & |[8]1,+1,0>_{6\,\bar 6} = \frac{1}{2\sqrt{5}}(2uu\bar u\bar d+2uu\bar d\bar u-2ud\bar d\bar d-2du\bar d\bar d+ & \\ \nonumber
      &  -us\bar d\bar s-us\bar s\bar d-su\bar d\bar s-su\bar s\bar d) & 
\end{eqnarray}
\begin{eqnarray}
 & |[8]1,0,0>_{6\,\bar 6} = \frac{1}{2\sqrt{10}}(4uu\bar u\bar u-4dd\bar d\bar d-ds\bar d\bar s-us\bar u\bar s+ & \\ \nonumber
      &  -ds\bar s\bar d-us\bar s\bar u-sd\bar d\bar s-su\bar u\bar s-sd\bar s\bar d-su\bar s\bar u) & 
\end{eqnarray}
\begin{eqnarray}
& |[8]1,-1,0>_{6\,\bar 6} = \frac{1}{2\sqrt{5}}(2du\bar u\bar u+2ud\bar u\bar u-2dd\bar u\bar d-2dd\bar d\bar u+ & \\ \nonumber
      &  -ds\bar u\bar s-ds\bar s\bar u-sd\bar u\bar s-sd\bar s\bar u) & 
\end{eqnarray}
\begin{eqnarray}
 & |[8]\frac{1}{2},+\frac{1}{2},+1>_{6\,\bar 6} = \frac{1}{2\sqrt{5}}(2uu\bar u\bar s+2uu\bar s\bar u-2us\bar s\bar s-2su\bar s\bar s+ & \\ \nonumber
      & -ud\bar d\bar s-ud\bar s\bar d-du\bar d\bar s-du\bar s\bar d ) & 
\end{eqnarray}
\begin{eqnarray}
 & |[8]\frac{1}{2},-\frac{1}{2},+1>_{6\,\bar 6} = \frac{1}{2\sqrt{5}}(-2ds\bar s\bar s-2sd\bar s\bar s-2dd\bar s\bar d-2dd\bar d\bar s+ & \\ \nonumber
      & +ud\bar u\bar s+du\bar u\bar s+ud\bar s\bar u+du\bar s\bar u ) & 
\end{eqnarray}
\begin{eqnarray}
 & |[8]\frac{1}{2},+\frac{1}{2},-1>_{6\,\bar 6} = \frac{1}{2\sqrt{5}}(-2sd\bar d\bar d-2ds\bar d\bar d-2ss\bar d\bar s-2ss\bar s\bar d+ & \\ \nonumber
      & +su\bar u\bar d+us\bar u\bar d+su\bar d\bar u+us\bar d\bar u ) & 
\end{eqnarray}
\begin{eqnarray}
 & |[8]\frac{1}{2},-\frac{1}{2},-1>_{6\,\bar 6} = \frac{1}{2\sqrt{5}}(2us\bar u\bar u+2su\bar u\bar u-2ss\bar u\bar s-2ss\bar s\bar u+ & \\ \nonumber
      & -sd\bar u\bar d-sd\bar d\bar u-ds\bar u\bar d-ds\bar d\bar u ) & 
\end{eqnarray}
\begin{eqnarray}
 & |[8]0,0,0>_{6\,\bar 6}  =  \frac{1}{2\sqrt{30}}(4uu\bar u\bar u+4dd\bar d\bar d-8ss\bar s\bar s+su\bar s\bar u+su\bar u\bar s+us\bar s\bar u+  & \\  \nonumber
 & +us\bar u\bar s-ds\bar d\bar s-ds\bar s\bar d-sd\bar d\bar s-sd\bar s\bar d-2ud\bar u\bar d-2ud\bar d\bar u-2du\bar u\bar d-2du\bar d\bar u) &
\end{eqnarray}
\begin{eqnarray}
 & |[10]\frac{3}{2},+\frac{3}{2},+1>_{6\,3} = \frac{1}{\sqrt{2}}(uu\bar d\bar s-uu\bar s\bar d) & 
\end{eqnarray}
\begin{eqnarray}
 & |[10]\frac{3}{2},+\frac{1}{2},+1>_{6\,3} = \frac{1}{\sqrt{6}}(uu\bar u\bar s-uu\bar s\bar u+du\bar d\bar s-du\bar s\bar d+ & \\ \nonumber
 & +ud\bar d\bar s-ud\bar s\bar d) &
\end{eqnarray}
\begin{eqnarray}
 & |[10]\frac{3}{2},-\frac{1}{2},+1>_{6\,3} = \frac{1}{\sqrt{6}}(dd\bar d\bar s-dd\bar s\bar d+du\bar u\bar s-du\bar s\bar u+ & \\ \nonumber
 & +ud\bar u\bar s-ud\bar s\bar u) & 
\end{eqnarray}  
\begin{eqnarray}
 & |[10]\frac{3}{2},-\frac{3}{2},+1>_{6\,3} = \frac{1}{\sqrt{2}}(dd\bar u\bar s-dd\bar s\bar u) & 
\end{eqnarray}
\begin{eqnarray}
 & |[10]1,+1,0>_{6\,3} = \frac{1}{\sqrt{6}}(uu\bar d\bar u-uu\bar u\bar d+us\bar d\bar s-us\bar s\bar d+ & \\ \nonumber
 & +su\bar d\bar s-su\bar s\bar d) &
\end{eqnarray}
\begin{eqnarray}
 & |[10]1,0,0>_{6\,3} = \frac{1}{2\sqrt{3}}(-du\bar u\bar d-ud\bar u\bar d+du\bar d\bar u+ud\bar d\bar u+ & \\ \nonumber
      &  -ds\bar s\bar d-us\bar s\bar u+ds\bar d\bar s+us\bar u\bar s+sd\bar d\bar s+su\bar u\bar s-sd\bar s\bar d-su\bar s\bar u) & 
\end{eqnarray}
\begin{eqnarray}
& |[10]1,-1,0>_{6\,3} = \frac{1}{\sqrt{6}}(dd\bar d\bar u-dd\bar u\bar d+ds\bar u\bar s-ds\bar s\bar u+ & \\ \nonumber
 & +sd\bar u\bar s-sd\bar s\bar u) &
\end{eqnarray}
\begin{eqnarray}
 & |[10]\frac{1}{2},+\frac{1}{2},-1>_{6\,3} = \frac{1}{\sqrt{6}}(us\bar d\bar u-us\bar u\bar d+ss\bar d\bar s-ss\bar s\bar d+ & \\ \nonumber
 & +su\bar d\bar u-su\bar u\bar d) &
\end{eqnarray}
\begin{eqnarray}
 & |[10]\frac{1}{2},-\frac{1}{2},-1>_{6\,3} = \frac{1}{\sqrt{6}}(ds\bar d\bar u-ds\bar u\bar d+ss\bar u\bar s-ss\bar s\bar u+ & \\ \nonumber
 & +sd\bar d\bar u-sd\bar u\bar d) &
\end{eqnarray}
\begin{eqnarray}
 & |[10]0,0,-2>_{6\,3}  =  \frac{1}{\sqrt{2}}(ss\bar d\bar u-ss\bar u\bar d)  & 
\end{eqnarray}
\begin{eqnarray}
 & |[\overline{10}]\frac{3}{2},+\frac{3}{2},-1>_{\bar 3\,\bar 6} = \frac{1}{\sqrt{2}}(us\bar d\bar d-su\bar d\bar d) & 
\end{eqnarray}
\begin{eqnarray}
 & |[\overline{10}]\frac{3}{2},+\frac{1}{2},-1>_{\bar 3\,\bar 6} = \frac{1}{\sqrt{6}}(ds\bar d\bar d-sd\bar d\bar d+us\bar u\bar d-su\bar u\bar d+ & \\ \nonumber
 & +us\bar d\bar u-su\bar d\bar u) &
\end{eqnarray}
\begin{eqnarray}
 & |[\overline{10}]\frac{3}{2},-\frac{1}{2},-1>_{\bar 3\,\bar 6} = \frac{1}{\sqrt{6}}(ds\bar u\bar d-sd\bar u\bar d+ds\bar d\bar u-sd\bar d\bar u+ & \\ \nonumber
 & +us\bar u\bar u-su\bar u\bar u) & 
\end{eqnarray}  
\begin{eqnarray}
 & |[\overline{10}]\frac{3}{2},-\frac{3}{2},-1>_{\bar 3\,\bar 6} = \frac{1}{\sqrt{2}}(ds\bar u\bar u-sd\bar u\bar u) & 
\end{eqnarray}
\begin{eqnarray}
 & |[\overline{10}]1,+1,0>_{\bar 3\,\bar 6} = \frac{1}{\sqrt{6}}(du\bar d\bar d-ud\bar d\bar d+us\bar d\bar s+us\bar s\bar d+ & \\ \nonumber
 & -su\bar d\bar s-su\bar s\bar d) &
\end{eqnarray}
\begin{eqnarray}
 & |[\overline{10}]1,0,0>_{\bar 3\,\bar 6} = \frac{1}{2\sqrt{3}}(du\bar u\bar d-ud\bar u\bar d+du\bar d\bar u-ud\bar d\bar u+ & \\ \nonumber
      &  +ds\bar s\bar d+us\bar s\bar u+ds\bar d\bar s+us\bar u\bar s-sd\bar d\bar s-su\bar u\bar s-sd\bar s\bar d-su\bar s\bar u) & 
\end{eqnarray}
\begin{eqnarray}
& |[\overline{10}]1,-1,0>_{\bar 3\,\bar 6} = \frac{1}{\sqrt{6}}(du\bar u\bar u-ud\bar u\bar u+ds\bar u\bar s+ds\bar s\bar u+ & \\ \nonumber
 & -sd\bar u\bar s-sd\bar s\bar u) &
\end{eqnarray}
\begin{eqnarray}
 & |[\overline{10}]\frac{1}{2},+\frac{1}{2},+1>_{\bar 3\,\bar 6} = \frac{1}{\sqrt{6}}(us\bar s\bar s-su\bar s\bar s+du\bar d\bar s-ud\bar d\bar s+ & \\ \nonumber
 & +du\bar s\bar d-ud\bar s\bar d) &
\end{eqnarray}
\begin{eqnarray}
 & |[\overline{10}]\frac{1}{2},-\frac{1}{2},+1>_{\bar 3\,\bar 6} = \frac{1}{\sqrt{6}}(ds\bar s\bar s-sd\bar s\bar s+du\bar u\bar s-ud\bar u\bar s+ & \\ \nonumber
 & +du\bar s\bar u-ud\bar s\bar u) &
\end{eqnarray}
\begin{eqnarray}
 & |[\overline{10}]0,0,+2>_{\bar 3\,\bar 6}  =  \frac{1}{\sqrt{2}}(du\bar s\bar s-ud\bar s\bar s)  & 
\end{eqnarray}
\begin{eqnarray}
 & |[27]2,+2,0>_{6\,\bar 6} = (uu\bar d\bar d) & 
\end{eqnarray}
\begin{eqnarray}
 & |[27]2,+1,0>_{6\,\bar 6}  =  \frac{1}{2}(du\bar d\bar d+ud\bar d\bar d+uu\bar u\bar d+uu\bar d\bar u)  & 
\end{eqnarray}
\begin{eqnarray}
& |[27]2,0,0>_{6\,\bar 6} = \frac{1}{\sqrt{6}}(uu\bar u\bar u+ud\bar u\bar d+dd\bar d\bar d+du\bar d\bar u+ & \\ \nonumber
      &  +du\bar u\bar d+ud\bar d\bar u) & 
\end{eqnarray}
\begin{eqnarray}
 & |[27]2,-1,0>_{6\,\bar 6}  =  \frac{1}{2}(du\bar u\bar u+ud\bar u\bar u+dd\bar u\bar d+dd\bar d\bar u)  & 
\end{eqnarray}
\begin{eqnarray}
 & |[27]2,-2,0>_{6\,\bar 6} = (dd\bar u\bar u) & 
\end{eqnarray}
\begin{eqnarray}
 & |[27]\frac{3}{2},+\frac{3}{2},+1>_{6\,\bar 6} = \frac{1}{\sqrt{2}}(uu\bar d\bar s+uu\bar s\bar d) & 
\end{eqnarray}
\begin{eqnarray}
 & |[27]\frac{3}{2},+\frac{1}{2},+1>_{6\,\bar 6} = \frac{1}{\sqrt{6}}(uu\bar u\bar s+uu\bar s\bar u+du\bar d\bar s+du\bar s\bar d+ & \\ \nonumber
 & +ud\bar d\bar s+ud\bar s\bar d) &
\end{eqnarray}
\begin{eqnarray}
 & |[27]\frac{3}{2},-\frac{1}{2},+1>_{6\,\bar 6} = \frac{1}{\sqrt{6}}(dd\bar d\bar s+dd\bar s\bar d+du\bar u\bar s+du\bar s\bar u+ & \\ \nonumber
 & +ud\bar u\bar s+ud\bar s\bar u) & 
\end{eqnarray}  
\begin{eqnarray}
 & |[27]\frac{3}{2},-\frac{3}{2},+1>_{6\,\bar 6} = \frac{1}{\sqrt{2}}(dd\bar u\bar s+dd\bar s\bar u) & 
\end{eqnarray}
\begin{eqnarray}
 & |[27]\frac{3}{2},+\frac{3}{2},-1>_{6\,\bar 6} = \frac{1}{\sqrt{2}}(us\bar d\bar d+su\bar d\bar d) & 
\end{eqnarray}
\begin{eqnarray}
 & |[27]\frac{3}{2},+\frac{1}{2},-1>_{6\,\bar 6} = \frac{1}{\sqrt{6}}(ds\bar d\bar d+sd\bar d\bar d+us\bar u\bar d+su\bar u\bar d+ & \\ \nonumber
 & +us\bar d\bar u+su\bar d\bar u) &
\end{eqnarray}
\begin{eqnarray}
 & |[27]\frac{3}{2},-\frac{1}{2},-1>_{6\,\bar 6} = \frac{1}{\sqrt{6}}(ds\bar u\bar d+sd\bar u\bar d+ds\bar d\bar u+sd\bar d\bar u+ & \\ \nonumber
 & +us\bar u\bar u+su\bar u\bar u) & 
\end{eqnarray}  
\begin{eqnarray}
 & |[27]\frac{3}{2},-\frac{3}{2},-1>_{6\,\bar 6} = \frac{1}{\sqrt{2}}(ds\bar u\bar u+sd\bar u\bar u) & 
\end{eqnarray}
\begin{eqnarray}
 & |[27]1,+1,+2>_{6\,\bar 6} = (uu\bar s\bar s) & 
\end{eqnarray}
\begin{eqnarray}
 & |[27]1,0,+2>_{6\,\bar 6}  =  \frac{1}{\sqrt{2}}(du\bar s\bar s+ud\bar s\bar s)  & 
\end{eqnarray}
\begin{eqnarray}
 & |[27]1,-1,+2>_{6\,\bar 6}  = (dd\bar s\bar s)  & 
\end{eqnarray}
\begin{eqnarray}
 & |[27]1,+1,-2>_{6\,\bar 6} = (ss\bar d\bar d) & 
\end{eqnarray}
\begin{eqnarray}
 & |[27]1,0,-2>_{6\,\bar 6}  =  \frac{1}{\sqrt{2}}(ss\bar u\bar d+ss\bar d\bar u)  & 
\end{eqnarray}
\begin{eqnarray}
 & |[27]1,-1,-2>_{6\,\bar 6}  = (ss\bar u\bar u)  & 
\end{eqnarray}
\begin{eqnarray}
 & |[27]1,+1,0>_{6\,\bar 6} = \frac{1}{2\sqrt{5}}(uu\bar u\bar d+uu\bar d\bar u-ud\bar d\bar d-du\bar d\bar d+ & \\ \nonumber
      &  +2us\bar d\bar s+2us\bar s\bar d+2su\bar d\bar s+2su\bar s\bar d) & 
\end{eqnarray}
\begin{eqnarray}
 & |[27]1,0,0>_{6\,\bar 6} = \frac{1}{\sqrt{10}}(uu\bar u\bar u-dd\bar d\bar d+ds\bar d\bar s+us\bar u\bar s+ & \\ \nonumber
      &  +ds\bar s\bar d+us\bar s\bar u+sd\bar d\bar s+su\bar u\bar s+sd\bar s\bar d+su\bar s\bar u) & 
\end{eqnarray}
\begin{eqnarray}
& |[27]1,-1,0>_{6\,\bar 6} = \frac{1}{2\sqrt{5}}(du\bar u\bar u+ud\bar u\bar u-dd\bar u\bar d-dd\bar d\bar u+ & \\ \nonumber
      &  +2ds\bar u\bar s+2ds\bar s\bar u+2sd\bar u\bar s+2sd\bar s\bar u) & 
\end{eqnarray}
\begin{eqnarray}
 & |[27]\frac{1}{2},+\frac{1}{2},+1>_{6\,\bar 6} = \frac{1}{\sqrt{30}}(2uu\bar u\bar s+2uu\bar s\bar u+3us\bar s\bar s+3su\bar s\bar s+ & \\ \nonumber
      & -ud\bar d\bar s-ud\bar s\bar d-du\bar d\bar s-du\bar s\bar d ) & 
\end{eqnarray}
\begin{eqnarray}
 & |[27]\frac{1}{2},-\frac{1}{2},+1>_{6\,\bar 6} = \frac{1}{\sqrt{30}}(-2dd\bar s\bar d-2dd\bar d\bar s+3ds\bar s\bar s+3sd\bar s\bar s+ & \\ \nonumber
      & +ud\bar u\bar s+du\bar u\bar s+ud\bar s\bar u+du\bar s\bar u ) & 
\end{eqnarray}
\begin{eqnarray}
 & |[27]\frac{1}{2},+\frac{1}{2},-1>_{6\,\bar 6} = \frac{1}{\sqrt{30}}(-2sd\bar d\bar d-2ds\bar d\bar d+3ss\bar d\bar s+3ss\bar s\bar d+ & \\ \nonumber
      & +su\bar u\bar d+us\bar u\bar d+su\bar d\bar u+us\bar d\bar u ) & 
\end{eqnarray}
\begin{eqnarray}
 & |[27]\frac{1}{2},-\frac{1}{2},-1>_{6\,\bar 6} = \frac{1}{\sqrt{30}}(2us\bar u\bar u+2su\bar u\bar u+3ss\bar u\bar s+3ss\bar s\bar u+ & \\ \nonumber
      & -sd\bar u\bar d-sd\bar d\bar u-ds\bar u\bar d-ds\bar d\bar u ) & 
\end{eqnarray}
\begin{eqnarray}
 & |[27]0,0,0>_{6\,\bar 6}  =  \frac{1}{2\sqrt{30}}(2uu\bar u\bar u+2dd\bar d\bar d+6ss\bar s\bar s+3su\bar s\bar u+3su\bar u\bar s+3us\bar s\bar u+  & \\  \nonumber
 & +3us\bar u\bar s-3ds\bar d\bar s-3ds\bar s\bar d-3sd\bar d\bar s-3sd\bar s\bar d-ud\bar u\bar d-ud\bar d\bar u-du\bar u\bar d-du\bar d\bar u) &
\end{eqnarray}

\subsection{\label{subapp:idealmixing}The $qq\bar q\bar q$ states in the \textquotedblleft ideal mixing\textquotedblright \, hypothesis}

In the \textquotedblleft ideal mixing\textquotedblright \, hypothesis, the flavor states of the tetraquarks are a superposition of the states written in section (\ref{subapp:statisapore})
in such a way to have defined strange quark and antiquark numbers. The notation used for the ideally mixed states is $|\text{strange\;quark\;number},I,I_{3},Y>_{R'\,R''}$.
Clearly the only states that can be mixed are those with the same quantum numbers (i.e. same isospin and same hypercharge). Only the mixed states are written below.

\begin{eqnarray}
 & |n_{S}=2,0,0,0>_{\bar 3\,3}=\frac{1}{\sqrt{3}}(\sqrt{2}|[8]0,0,0>_{\bar 3\,3}+|[1]0,0,0>_{\bar 3\,3})= & \nonumber \\
 & =\frac{1}{2\sqrt{2}}(-su\bar s\bar u+su\bar u\bar s+us\bar s\bar u-us\bar u\bar s+ds\bar d\bar s-ds\bar s\bar d-sd\bar d\bar s+sd\bar s\bar d) &  
\end{eqnarray}
\begin{eqnarray}
 & |n_{S}=0,0,0,0>_{\bar 3\,3}=\frac{1}{\sqrt{3}}(|[8]0,0,0>_{\bar 3\,3}-\sqrt{2}|[1]0,0,0>_{\bar 3\,3})= & \nonumber \\
 & =\frac{1}{2}(ud\bar u\bar d-ud\bar d\bar u-du\bar u\bar d+du\bar d\bar u) & 
\end{eqnarray}

\begin{eqnarray}
 & |n_{S}=2,1,+1,0>_{\bar 3\,\bar 6}=\frac{1}{\sqrt{3}}(\sqrt{2}|[\overline{10}]1,+1,0>_{\bar 3\,\bar 6}+|[8]1,+1,0>_{\bar 3\,\bar 6})= & \nonumber \\
 & =\frac{1}{2}(us\bar d\bar s+us\bar s\bar d-su\bar d\bar s-su\bar s\bar d) &  
\end{eqnarray}
\begin{equation}
|n_{S}=2,1,0,0>_{\bar 3\,\bar 6}= \frac{1}{2\sqrt{2}}(ds\bar d\bar s+us\bar u\bar s+ds\bar s\bar d+us\bar s\bar u-sd\bar d\bar s-su\bar u\bar s-sd\bar s\bar d-su\bar s\bar u)  
\end{equation} 
\begin{equation}
|n_{S}=2,1,-1,0>_{\bar 3\,\bar 6}=\frac{1}{2}(ds\bar u\bar s+ds\bar s\bar u-sd\bar u\bar s-sd\bar s\bar u)
\end{equation}
\begin{eqnarray}
 & |n_{S}=0,1,+1,0>_{\bar 3\,\bar 6}=\frac{1}{\sqrt{3}}(|[\overline{10}]1,+1,0>_{\bar 3\,\bar 6}-\sqrt{2}|[8]0,0,0>_{\bar 3\,\bar 6})= & \nonumber \\
 & =\frac{1}{\sqrt{2}}(du\bar d\bar d-ud\bar d\bar d) & 
\end{eqnarray}
\begin{equation}
|n_{S}=2,1,0,0>_{\bar 3\,\bar 6}= \frac{1}{2}(du\bar u\bar d+du\bar d\bar u-ud\bar u\bar d-ud\bar d\bar u)  
\end{equation}
\begin{equation}
|n_{S}=0,1,-1,0>_{\bar 3\,\bar 6}=\frac{1}{\sqrt{2}}(du\bar u\bar u-ud\bar u\bar u)
\end{equation}
\begin{eqnarray}
 & |n_{S}=3,\frac{1}{2},+\frac{1}{2},+1>_{\bar 3\,\bar 6}=\frac{1}{\sqrt{3}}(|[\overline{10}]\frac{1}{2},+\frac{1}{2},+1>_{\bar 3\,\bar 6}+\sqrt{2}|[8]\frac{1}{2},+\frac{1}{2},+1>_{\bar 3\,\bar 6})= & \nonumber \\
 & =\frac{1}{\sqrt{2}}(us\bar s\bar s-su\bar s\bar s) &  
\end{eqnarray}
\begin{equation}
|n_{S}=3,\frac{1}{2},-\frac{1}{2},+1>_{\bar 3\,\bar 6}= \frac{1}{\sqrt{2}}(ds\bar s\bar s-sd\bar s\bar s)  
\end{equation}
\begin{eqnarray}
 & |n_{S}=1,\frac{1}{2},+\frac{1}{2},+1>_{\bar 3\,\bar 6}=\frac{1}{\sqrt{3}}(-\sqrt{2}|[\overline{10}]\frac{1}{2},+\frac{1}{2},+1>_{\bar 3\,\bar 6}+|[8]\frac{1}{2},+\frac{1}{2},+1>_{\bar 3\,\bar 6})= & \nonumber \\
 & =\frac{1}{2}(ud\bar d\bar s-du\bar d\bar s+ud\bar s\bar d-du\bar s\bar d) &  
\end{eqnarray}
\begin{equation}
|n_{S}=1,\frac{1}{2},-\frac{1}{2},+1>_{\bar 3\,\bar 6}= \frac{1}{2}(ud\bar u\bar s-du\bar u\bar s+ud\bar s\bar u-du\bar s\bar u)  
\end{equation}

\begin{eqnarray}
 & |n_{S}=2,1,+1,0>_{6\,3}=\frac{1}{\sqrt{3}}(-\sqrt{2}|[10]1,+1,0>_{6\,3}+|[8]1,+1,0>_{6\,3})= & \nonumber \\
 & =\frac{1}{2}(us\bar s\bar d-us\bar d\bar s-su\bar d\bar s+su\bar s\bar d) & 
\end{eqnarray}
\begin{equation}
|n_{S}=2,1,0,0>_{6\,3}= \frac{1}{2\sqrt{2}}(-ds\bar d\bar s-us\bar u\bar s+ds\bar s\bar d+us\bar s\bar u-sd\bar d\bar s-su\bar u\bar s+sd\bar s\bar d+su\bar s\bar u)  
\end{equation}
\begin{equation}
|n_{S}=2,1,-1,0>_{6\,3}=\frac{1}{2}(-ds\bar u\bar s+ds\bar s\bar u-sd\bar u\bar s+sd\bar s\bar u)
\end{equation}
\begin{eqnarray}
 & |n_{S}=0,1,+1,0>_{6\,3}=\frac{1}{\sqrt{3}}(|[10]1,+1,0>_{6\,3}+\sqrt{2}|[8]1,+1,0>_{6\,3})= & \nonumber \\
 & =\frac{1}{\sqrt{2}}(uu\bar d\bar u-uu\bar u\bar d) & 
\end{eqnarray}
\begin{equation}
|n_{S}=0,1,0,0>_{6\,3}= \frac{1}{2}(-du\bar u\bar d+du\bar d\bar u-ud\bar u\bar d+ud\bar d\bar u)  
\end{equation}
\begin{equation}
|n_{S}=0,1,-1,0>_{6\,3}=\frac{1}{\sqrt{2}}(dd\bar d\bar u-dd\bar u\bar d)
\end{equation}
\begin{eqnarray}
 & |n_{S}=3,\frac{1}{2},+\frac{1}{2},-1>_{6\,3}=\frac{1}{\sqrt{3}}(|[10]\frac{1}{2},+\frac{1}{2},-1>_{6\,3}-\sqrt{2}|[8]\frac{1}{2},+\frac{1}{2},-1>_{6\,3})= & \nonumber \\
 & =\frac{1}{\sqrt{2}}(ss\bar d\bar s-ss\bar s\bar d) &  
\end{eqnarray}
\begin{equation}
|n_{S}=3,\frac{1}{2},-\frac{1}{2},-1>_{6\,3}= \frac{1}{\sqrt{2}}(ss\bar u\bar s-ss\bar s\bar u)  
\end{equation}
\begin{eqnarray}
 & |n_{S}=1,\frac{1}{2},+\frac{1}{2},-1>_{6\,3}=\frac{1}{\sqrt{3}}(\sqrt{2}|[10]\frac{1}{2},+\frac{1}{2},-1>_{6\,3}+|[8]\frac{1}{2},+\frac{1}{2},-1>_{6\,3})= & \nonumber \\
 & =\frac{1}{2}(su\bar d\bar u-su\bar u\bar d+us\bar d\bar u-us\bar u\bar d) &  
\end{eqnarray}
\begin{equation}
|n_{S}=1,\frac{1}{2},-\frac{1}{2},-1>_{6\,3}= \frac{1}{2}(sd\bar d\bar u-sd\bar u\bar d+ds\bar d\bar u-ds\bar u\bar d) 
\end{equation}

\begin{eqnarray}
 & |n_{S}=2,1,+1,0>_{6\,\bar 6}=\frac{1}{\sqrt{5}}(2|[27]1,+1,0>_{6\,\bar 6}-|[8]1,+1,0>_{6\,\bar 6})= & \nonumber \\
 & =\frac{1}{2}(us\bar s\bar d+us\bar d\bar s+su\bar d\bar s+su\bar s\bar d) &  
\end{eqnarray}
\begin{equation}
|n_{S}=2,1,0,0>_{6\,\bar 6}= \frac{1}{2\sqrt{2}}(ds\bar d\bar s+us\bar u\bar s+ds\bar s\bar d+us\bar s\bar u+sd\bar d\bar s+su\bar u\bar s+sd\bar s\bar d+su\bar s\bar u)  
\end{equation}
\begin{equation}
|n_{S}=2,1,-1,0>_{6\,\bar 6}=\frac{1}{2}(ds\bar u\bar s+ds\bar s\bar u+sd\bar u\bar s+sd\bar s\bar u)
\end{equation}
\begin{eqnarray}
 & |n_{S}=0,1,+1,0>_{6\,\bar 6}=\frac{1}{\sqrt{5}}(-|[27]1,+1,0>_{6\,\bar 6}-2|[8]1,+1,0>_{6\,\bar 6})= & \nonumber \\
 & =\frac{1}{2}(ud\bar d\bar d+du\bar d\bar d-uu\bar u\bar d-uu\bar d\bar u) & 
\end{eqnarray}
\begin{equation}
|n_{S}=0,1,0,0>_{6\,\bar 6}= \frac{1}{\sqrt{2}}(dd\bar d\bar d-uu\bar u\bar u)  
\end{equation} 
\begin{equation}
|n_{S}=0,1,-1,0>_{6\,\bar 6}=\frac{1}{2}(dd\bar u\bar d+dd\bar d\bar u-du\bar u\bar u-ud\bar u\bar u)
\end{equation}
\begin{eqnarray}
 & |n_{S}=3,\frac{1}{2},+\frac{1}{2},-1>_{6\,\bar 6}=\frac{1}{\sqrt{5}}(\sqrt{3}|[27]\frac{1}{2},+\frac{1}{2},-1>_{6\,\bar 6}-\sqrt{2}|[8]\frac{1}{2},+\frac{1}{2},-1>_{6\,\bar 6})= & \nonumber \\
 & =\frac{1}{\sqrt{2}}(ss\bar d\bar s+ss\bar s\bar d) &  
\end{eqnarray}
\begin{equation}
|n_{S}=3,\frac{1}{2},-\frac{1}{2},-1>_{6\,\bar 6}= \frac{1}{\sqrt{2}}(ss\bar u\bar s+ss\bar s\bar u)  
\end{equation}
\begin{eqnarray}
 & |n_{S}=1,\frac{1}{2},+\frac{1}{2},-1>_{6\,\bar 6}=\frac{1}{\sqrt{5}}(\sqrt{2}|[27]\frac{1}{2},+\frac{1}{2},-1>_{6\,\bar 6}+\sqrt{3}|[8]\frac{1}{2},+\frac{1}{2},-1>_{6\,\bar 6})= & \nonumber \\
 & =\frac{1}{2\sqrt{3}}(su\bar d\bar u+su\bar u\bar d+us\bar d\bar u+us\bar u\bar d-2sd\bar d\bar d-2ds\bar d\bar d) &  
\end{eqnarray}
\begin{equation}
|n_{S}=1,\frac{1}{2},-\frac{1}{2},-1>_{6\,\bar 6}= \frac{1}{2\sqrt{3}}(2us\bar u\bar u+2su\bar u\bar u-sd\bar d\bar u-sd\bar u\bar d-ds\bar d\bar u-ds\bar u\bar d) 
\end{equation}
\begin{eqnarray}
 & |n_{S}=3,\frac{1}{2},+\frac{1}{2},+1>_{6\,\bar 6}=\frac{1}{\sqrt{5}}(\sqrt{3}|[27]\frac{1}{2},+\frac{1}{2},+1>_{6\,\bar 6}-\sqrt{2}|[8]\frac{1}{2},+\frac{1}{2},+1>_{6\,\bar 6})= & \nonumber \\
 & =\frac{1}{\sqrt{2}}(us\bar s\bar s+su\bar s\bar s) &  
\end{eqnarray}
\begin{equation}
|n_{S}=3,\frac{1}{2},-\frac{1}{2},+1>_{6\,\bar 6}= \frac{1}{\sqrt{2}}(ds\bar s\bar s+sd\bar s\bar s)  
\end{equation}
\begin{eqnarray}
 & |n_{S}=1,\frac{1}{2},+\frac{1}{2},+1>_{6\,\bar 6}=\frac{1}{\sqrt{5}}(\sqrt{2}|[27]\frac{1}{2},+\frac{1}{2},+1>_{6\,\bar 6}+\sqrt{3}|[8]\frac{1}{2},+\frac{1}{2},+1>_{6\,\bar 6})= & \nonumber \\
 & =\frac{1}{2\sqrt{3}}(2uu\bar u\bar s+2uu\bar s\bar u-ud\bar d\bar s-du\bar d\bar s-ud\bar s\bar d-du\bar s\bar d) &  
\end{eqnarray}
\begin{equation}
|n_{S}=1,\frac{1}{2},-\frac{1}{2},+1>_{6\,\bar 6}= \frac{1}{2\sqrt{3}}(ud\bar u\bar s+du\bar u\bar s+ud\bar s\bar u+du\bar s\bar u-2dd\bar d\bar s-2dd\bar s\bar d)  
\end{equation}
\begin{eqnarray}
 & |n_{S}=4,0,0,0>_{6\,\bar 6}=\frac{\sqrt{3}}{\sqrt{10}}|[27]0,0,0>_{6\,\bar 6}-\frac{2\sqrt{2}}{\sqrt{15}}|[8]0,0,0>_{6\,\bar 6}+\frac{1}{\sqrt{6}}|[1]0,0,0>_{6\,\bar 6}= & \nonumber \\
 & =ss\bar s\bar s                      & 
\end{eqnarray}
\begin{eqnarray}
 & |n_{S}=2,0,0,0>_{6\,\bar 6}=-\frac{\sqrt{3}}{\sqrt{5}}|[27]0,0,0>_{6\,\bar 6}-\frac{1}{\sqrt{15}}|[8]0,0,0>_{6\,\bar 6}+\frac{1}{\sqrt{3}}|[1]0,0,0>_{6\,\bar 6}= & \nonumber \\
 & =\frac{1}{2\sqrt{2}}(-us\bar u\bar s-us\bar s\bar u-su\bar u\bar s-su\bar s\bar u+ds\bar d\bar s+ds\bar s\bar d+sd\bar d\bar s+sd\bar s\bar d) &  
\end{eqnarray}
\begin{eqnarray}
 & |n_{S}=0,0,0,0>_{6\,\bar 6}=-\frac{1}{\sqrt{10}}|[27]0,0,0>_{6\,\bar 6}-\frac{\sqrt{2}}{\sqrt{5}}|[8]0,0,0>_{6\,\bar 6}-\frac{1}{\sqrt{2}}|[1]0,0,0>_{6\,\bar 6}= & \nonumber \\
 & =\frac{1}{2\sqrt{3}}(ud\bar u\bar d+du\bar u\bar d+ud\bar d\bar u+du\bar d\bar u-2uu\bar u\bar u-2dd\bar d\bar d) &  
\end{eqnarray}
\section{\label{app:statispin} The tetraquark spin states}

In this appendix we write the spin states of the tetraquarks in terms of the spins of the single quarks and antiquarks.
These states are calculated by using the SU(2) Clebsch-Gordan coefficients. 

A generic spin state is expressed by $|[R]\;S,\;S_{z}>$, where $[R]$ indicates the SU(2)$_{S}$ representation, $S$ the spin quantum number and $S_{z}$ its third component.
The single quark states are written in short as: $\uparrow \equiv |[2]\;\frac{1}{2}\;+\frac{1}{2}> $ and $\downarrow \equiv |[2]\;\frac{1}{2}\;-\frac{1}{2}>$.

In this appendix the tetraquark states are written in the $(qq)(\bar q\bar q)$ configuration; thus, in addition to the representation $[R]$ to which the state belongs, we also show the representations $[R']$ and $[R'']$ respectively of the two quarks and the two antiquarks.
In short a tetraquark state will be written as $|[R]\;S,\;S_{z}>_{R'\;R''}$.

\subsection{\label{subapp:statispin}Tetraquark spin states in the $(qq)(\bar q\bar q)$ configuration}

\begin{equation}
|[1]\;0\;0>_{1\,1}=\frac{1}{2}(\uparrow \downarrow \uparrow \downarrow +\downarrow \uparrow \downarrow \uparrow -\uparrow \downarrow \downarrow \uparrow -\downarrow \uparrow \uparrow \downarrow )
\end{equation}

\begin{equation}
|[1]\;0\;0>_{3\,3}=\frac{1}{2\sqrt{3}}(2\uparrow \uparrow \downarrow \downarrow +2\downarrow \downarrow \uparrow \uparrow +\uparrow \downarrow \uparrow \downarrow +\downarrow \uparrow \downarrow \uparrow +\uparrow \downarrow \downarrow \uparrow +\downarrow \uparrow \uparrow \downarrow )
\end{equation}

\begin{equation}
|[3]\;1\;+1>_{3\,1}=\frac{1}{\sqrt{2}}(\uparrow \uparrow \uparrow \downarrow -\uparrow \uparrow \downarrow \uparrow  )
\end{equation}
\begin{equation}
|[3]\;1\;0>_{3\,1}=\frac{1}{2}(\uparrow \downarrow \uparrow \downarrow +\downarrow \uparrow \uparrow \downarrow -\uparrow \downarrow \downarrow \uparrow -\downarrow \uparrow \downarrow \uparrow  )
\end{equation}
\begin{equation}
|[3]\;1\;-1>_{3\,1}=\frac{1}{\sqrt{2}}(\downarrow \downarrow \uparrow \downarrow -\downarrow \downarrow \downarrow \uparrow  )
\end{equation}

\begin{equation}
|[3]\;1\;+1>_{1\,3}=\frac{1}{\sqrt{2}}(\uparrow \downarrow \uparrow \uparrow -\downarrow \uparrow \uparrow \uparrow  )
\end{equation}
\begin{equation}
|[3]\;1\;0>_{1\,3}=\frac{1}{2}(\uparrow \downarrow \uparrow \downarrow +\uparrow \downarrow \downarrow \uparrow -\downarrow \uparrow \uparrow \downarrow -\downarrow \uparrow \downarrow \uparrow  )
\end{equation}
\begin{equation}
|[3]\;1\;-1>_{1\,3}=\frac{1}{\sqrt{2}}(\uparrow \downarrow \downarrow \downarrow -\downarrow \uparrow \downarrow \downarrow  )
\end{equation}

\begin{equation}
|[3]\;1\;+1>_{3\,3}=\frac{1}{2}(\uparrow \uparrow \uparrow \downarrow +\uparrow \uparrow \downarrow \uparrow -\uparrow \downarrow \uparrow \uparrow -\downarrow \uparrow \uparrow \uparrow  )
\end{equation}
\begin{equation}
|[3]\;1\;0>_{3\,3}=\frac{1}{\sqrt{2}}(\uparrow \uparrow \downarrow \downarrow -\downarrow \downarrow \uparrow \uparrow )
\end{equation}
\begin{equation}
|[3]\;1\;-1>_{3\,3}=\frac{1}{2}(-\downarrow \downarrow \uparrow \downarrow -\downarrow \downarrow \downarrow \uparrow +\uparrow \downarrow \downarrow \downarrow +\downarrow \uparrow \downarrow \downarrow  )
\end{equation}

\begin{equation}
|[5]\;2\;+2>_{3\,3}=\uparrow \uparrow \uparrow \uparrow 
\end{equation}
\begin{equation}
|[5]\;2\;+1>_{3\,3}=\frac{1}{2}(\uparrow \uparrow \uparrow \downarrow +\uparrow \uparrow \downarrow \uparrow +\uparrow \downarrow \uparrow \uparrow +\downarrow \uparrow \uparrow \uparrow ) 
\end{equation}
\begin{equation}
|[5]\;2\;0>_{3\,3}=\frac{1}{\sqrt{6}}(\uparrow \uparrow \downarrow \downarrow +\downarrow \downarrow \uparrow \uparrow +\uparrow \downarrow \uparrow \downarrow +\uparrow \downarrow \downarrow \uparrow +\downarrow \uparrow \uparrow \downarrow +\downarrow \uparrow \downarrow \uparrow ) 
\end{equation}
\begin{equation}
|[5]\;2\;-1>_{3\,3}=\frac{1}{2}(\downarrow \downarrow \uparrow \downarrow +\downarrow \downarrow \downarrow \uparrow +\uparrow \downarrow \downarrow \downarrow +\downarrow \uparrow \downarrow \downarrow  ) 
\end{equation}
\begin{equation}
|[5]\;2\;-2>_{3\,3}=\downarrow \downarrow \downarrow \downarrow  
\end{equation}
\end{appendix}

\end{document}